    \documentclass[12pt]{article} % use larger type; default would be 10pt
    
    \usepackage[utf8]{inputenc} % set input encoding (not needed with XeLaTeX)
    
    \usepackage{geometry,color} % to change the page dimensions
    \usepackage{graphicx,amsmath,yfonts,amssymb,hyperref} % support the \includegraphics command and options
   
    \usepackage{natbib}
    
    \usepackage{booktabs} % for much better looking tables
    \usepackage{array} % for better arrays (eg matrices) in maths
    \usepackage{paralist} % very flexible & customisable lists (eg. enumerate/itemize, etc.)
    \usepackage{verbatim} % adds environment for commenting out blocks of text & for better verbatim
    \usepackage{subfig} % make it possible to include more than one captioned figure/table in a single float
    \usepackage{fancyhdr} % This should be set AFTER setting up the page geometry
\newcommand{\xL}{\chi}

    \usepackage{sectsty}
    \usepackage{amsmath,amsthm} % Required for \varPsi below
    
    \usepackage{tikz}
    \usetikzlibrary[positioning,calc,decorations.pathreplacing]
    
    \newtheorem{lemma}{Lemma}
    \newtheorem{theorem}{Theorem}

    \usepackage[onehalfspacing]{setspace}
    
    \usepackage[nottoc,notlof,notlot]{tocbibind} % Put the bibliography in the ToC
    \usepackage[titles,subfigure]{tocloft} % Alter the style of the Table of Contents
    
\newtheorem{innercustomthm}{Assumption}
\newenvironment{customasp}[1]
  {\renewcommand\theinnercustomthm{#1}\innercustomthm}
  {\endinnercustomthm}
 
     \newtheorem{innercustomlem}{Lemma}
  \newenvironment{customlemma}[1]
  {\renewcommand\theinnercustomlem{#1}\innercustomlem}
  {\endinnercustomlem}
  
       \newtheorem{innercustomthem}{Theorem}
  \newenvironment{customthem}[1]
  {\renewcommand\theinnercustomthem{#1}\innercustomthem}
  {\endinnercustomthem}
    \title{Identification of Auction Models \\ Using Order Statistics\thanks{We thank Yonghong An, Emmanuel Guerre, Philip Haile, Yingyao Hu, Koen Jochmans, Brad Larsen, Arthur Lewbel, Eric Mbakop, Isabelle Perrigne, Daniel Quint, Xun Tang, Quang Vuong, Yi Xin, and seminar participants at the University of Calgary, TAMU, IIOC (2019), Shanghai Econometric Workshop (2019), Econometric Society Meetings (China 2019, European Winter 2020, World Congress 2020), and AEA (2020, 2021). We would also like
to thank the editor and three anonymous referees for their comments. JoonHwan Cho and Jiaqi Zou provided valuable research assistance. Luo acknowledges funding from the SSHRC Insight Grant. The usual disclaimer applies. Luo: Department of Economics, University of Toronto, 150 St.\ George St, Toronto, ON M5S 3G7. Email: yao.luo@utoronto.ca. Xiao: Department of Economics, Indiana University, 100 S.\ Woodlawn Ave., Bloomington, IN 47405. Email: rulixiao@iu.edu.}}
    \author{Yao Luo
    \and Ruli Xiao }
    \date{January 2023} % Activate to display a given date or no date (if empty),
\begin{document}
    
    \maketitle
    
    \vspace{-1cm}
    \begin{abstract}
Auction data often contain information on only the most competitive bids as opposed to all bids. The usual measurement error approaches to unobserved heterogeneity are inapplicable due to dependence among order statistics. We bridge this gap by providing a set of positive identification results. First, we show that symmetric auctions with discrete unobserved heterogeneity are identifiable using two consecutive order statistics and an instrument. Second, we extend the results to ascending auctions with unknown competition and unobserved heterogeneity. 
    \end{abstract}
    
    \vspace{0.05cm}
    
    %\noindent 
    
    \vspace{0.05cm}
    
          
    \section{Introduction} 
    
    Recent years have witnessed a rapid growth in the literature combining auction theory with econometric analysis to understand auction markets and inform policies. While the key elements in auction theory generally match well with the empirics, this may not be the case when there is auction-level unobserved heterogeneity, i.e., factors affecting bidder values that are common knowledge among bidders but unobserved by the analyst. Ignoring such unobserved heterogeneity (UH) would lead to erroneous estimates and misinformed policy conclusions. See, e.g., \citet{HK2019} for a survey. 
    
Measurement error approaches exploit the multiplicity of bids in each auction to account for UH. See, e.g., \citet{li2000conditionally}, \citet{K2011} and \citet{HMS2013}.\footnote{The earliest approach to allow for UH in auctions exploits an auxiliary variable that is monotone in UH. See, e.g., \citet{CGPV2003} and \citet{HHS2003}.} They require observed bids to be independent conditioning on UH. In practice, however, this conditional independence assumption may not hold due to auction format or data truncation. Often, we may only observe multiple order statistics or a subset of bids rather than all bids themselves. For instance, the highest bid is never observed in ascending auctions.\footnote{For instance, \citet{KL2014} observes the second, third, and fourth highest bids in ascending used-car auctions. \citet{FL2017} uses the second and third order statistics of bids in eBay auctions for used iPhones.} Moreover, the low/high bids are much easier to access in many settings.\footnote{For instance, three apparent low bids in all auctions since 1999 are one-click downloadable on the California DOT website. The Washington State DOT archives bid opening results of six months or older online, but only for three apparent low bids. \citet{allen2019resolving} studies FDIC auction data, which contain the bid and the identity of the individuals associated with the winning and second-highest bids. Unfortunately, the data miss the names associated with all other bids. U.S. Forest Service timber auctions only record at most the top 12 bids regardless of the number of bidders.} Naturally, these order statistics are dependent even though the bids themselves are independent, thereby rendering the conventional measurement error approaches inapplicable.

%\footnote{Bid opening results can be found at https://www.wsdot.wa.gov/biz/contaa/Bidopening.}

%\footnote{See \href{https://www.dropbox.com/s/we2rfiv7llio9km/summary_order_stat.pdf?dl=0}{Online Appendix} for more examples of auction data containing only order statistics.}

This paper provides a set of positive results on the identification of auction models with \textit{discrete} UH using multiple order statistics of bids.\footnote{In the classical measurement error setting, \citet{AH2002} suggests ``there may be sufficient structure to identify the model from only two order statistics... However, we have not obtained such a result.'' Recently, \citet{luo2022two} obtains this result. Another companion paper, \citet{luo2020order}, considers nonseparable continuous UH using three consecutive order statistics.} Rather than circumventing the correlation among order statistics, we propose new identification strategies that exploit this very correlation structure with the same number of bids as the conventional measurement error approaches. Our results provide new perspectives into the identification of auction models with UH when the conditional independence assumption fails. Moreover, this opens a new window to explore the use of statistical/model structure to restore identification.

We consider a common form of incomplete bid data: \textit{consecutive} order statistics of bids. All the examples mentioned above have this form. Despite their correlation, we derive that the joint distribution of consecutive order statistics has a semi-separable structure with a type-independent indicator function capturing the correlation. Based on this finding, we show that two consecutive order statistics and an instrument or three consecutive ones identify independent private value (IPV) auction models with nonseparable finite UH. 

Our first result considers a benchmark case with symmetric bidders and observed competition (i.e., known number of bidders). We allow the cardinality of UH's support to be unknown and show its identification using two consecutive ordered bids. Given this cardinality, we first identify the component bid distributions and then link them to the model primitives --- value distributions. 

We propose a novel multi-step procedure using two consecutive order statistics and an instrument. First, exploiting the joint probability of observed bids and their being in two ordered intervals, we turn the semi-separable joint probability into an eigendecomposition structure and identify the distribution of an order statistic up to scale in each interval. Second, using its relationship with order statistics, we identify the parent bid distribution up to (different) scales in these intervals. Third, we pin down the scales exploiting the continuity of the component bid density functions. In ascending auctions, the bid distributions equal the value distributions. In first-price auctions, we follow \citet{GPV2000} to recover the value distributions using the bid distribution and the observed competition. When an instrument is difficult to find, our method extends to the scenario of three \textit{consecutive} order statistics. 

Our second result concerning ascending auctions allows competition as an additional dimension of UH, which is common with incomplete bid data. Missing information on competition leads to multiple sources of UH: auction state and number of bidders. This has rarely been dealt with in the auction literature. In ascending auctions, bidding strategies and bid distributions are independent of competition intensity. We extend \citet{S2004} by allowing for UH. On the one hand, we follow the paper to ``concentrate out'' unknown competition intensity. This transforms our problem into a mixture problem with just an unknown auction state and identifies the component bid distributions up to scales. On the other hand, we argue that \citet{S2004}'s identification of the scales at the limit is at odds with our setting. Instead, we propose a novel approach to pin down competition intensity and the scales by combining the joint and marginal distributions of the observed order statistics. Without UH, we identify the value distribution and the distribution of competition using two consecutive order statistics; with UH, we can do the same by using an additional instrument or order statistic. In contrast, \citet{S2004}'s approach cannot identify the distribution of competition under UH.

    \subsection*{Literature Review}
       
    There has been growing interest in using order statistics for identification in the auction literature. Many markets where only the transaction price is observed can be treated as auctions. \citet{AH2002} shows that the auction model with symmetric IPV is identifiable with the transaction price and the number of bidders. This exploits a one-to-one mapping between the distribution of an order statistic and its parent distribution. \citet{komarova2013new} shows that the asymmetric second‐price auction model is identifiable solely using the winner's identity and the transaction price.  \citet{GL2018} shows that the first-price auction model with symmetric IPV and unknown competition is identifiable solely using the transaction price and exploiting discontinuities in the density function of the winning bid due to changing competition. Our paper is the first to consider \textit{consecutive} order statistics for identification; it allows a general form of UH as opposed to unknown competition in \citet{GL2018}. In contrast to partial identification approaches (see, e.g., \citet{haile2003inference} and \citet{AGQ2013}), our approach leads to point identification and thus allows for the calculation of optimal reserve prices conditional on the realization of UH. 

    Our paper focuses on discrete nonseparable UH. The existing literature that allows for finite UH uses the eigendecomposition approach, as in \citet{HMS2013}. Identification is achieved using the joint distribution of at least three bidders per auction, following \citet{H2008}. This approach has been adapted to various settings. See, e.g., \citet{an2010estimating} for unknown competition, \citet{hortaccsu2018empirical} for auctions of multiple objects, and \citet{Luo2018} for multiple sources of UH such as the auction state and the number of bidders. Related to our first result, \citet{M2017} takes an alternative approach that attempts to restore the conditional independence structure needed in the usual approaches. In particular, he exploits the Markov property of order statistics and shows that the bidder's UH-specific distribution and the distribution of UH is point identified from (any) five order statistics. In contrast, we take advantage of the structure provided by consecutive order statistics and achieve identification using two consecutive order statistics and an instrument or three consecutive ones. 

Our extension contributes to the literature dealing with unobservability of competition to the analyst. This literature begins with \citet{laffont1995econometrics}, which treats unknown competition as an unknown parameter in first-price auctions. Recent contributions treat unknown competition as random (see, e.g., \citet{an2010estimating}, \citet{shneyerov2011identification}, and \citet{GL2018} for first-price auctions). Regarding ascending auctions, \citet{S2004} considers eBay auctions where the number of bidders is random and unobservable, showing that any two order statistics identify the symmetric IPV model without UH. \citet{KL2014} applies the identification results of \citet{S2004} to wholesale used-car auctions in Korea, for which only a small subset of bids is observable (due to the ascending auction format). \citet{FL2017} studies eBay auctions for used iPhones with auction-specific UH and an unknown number of bidders. The issue of correlated order statistics is circumvented using observed reserve prices, returning the identification problem to its standard form. Our paper extends this literature on unknown competition by introducing multiple sources of UH: auction state and the number of bidders.

Another literature considers continuous UH using the deconvolution approach, as in \citet{LV1998}, \citet{li2000conditionally}, and \citet{K2011}, among others.\footnote{\citet{chen2011nonlinear} provides an excellent survey of measurement error models.} Two random bids are required in each auction, and UH is restricted to have a separable structure on bidder valuation. \citet{HQT2019} achieves point identification using English auction models with additively separable UH, assuming piecewise real analytic density functions and using variations in the number of bidders across auctions. We focus on finite UH, which does not directly extend to the deconvolution approach. 
    
The remainder of this paper is organized as follows. Section 2 describes the auction models. Section 3 explains some useful properties of order statistics. Section 4 presents our main identification results.  Section 5 concludes. The Appendix contains longer proofs. 

%Section 5 extends the identification results for asymmetric auctions. Section 6 presents an application to USFS timber auctions. Section 7 concludes. 
    
    %presents the estimation strategy and Monte Carlo evidence of finite sample property. Section 6 applies the proposed method to study DOT auctions, and section 7
    
    \section{Auction Models} \label{model}
    
We consider symmetric standard independent private value (IPV) auction models. In practice, the auctioned items can be heterogeneous. While bidders observe such heterogeneity, the analyst  may not observe all auction characteristics, resulting in auction-level unobserved heterogeneity (UH). Hereafter, we refer to the realization of the auction-level UH as ``state.'' For the remainder of this section and Section \ref{OS}, we present results that condition on a realization of UH. In Section \ref{mainid}, we return to dealing with the distribution of UH.

Suppose that, for a given auction, the bidder values are i.i.d.\ draws from the same distribution $\Phi^k(v)$, where auction state $k$ is finite and discrete, i.e., $k=1,...,K$, and state $k$ is realized with probability $p_k>0$, and $\sum^K_{k=1}  p_k=1$.  Hereafter, we abstract from observable (to the analyst) characteristics for simplicity. Our results are applicable to models with both types of heterogeneity.

Suppose $n \geq 2$ symmetric bidders participate in an auction with a zero reserve price. Bidders are risk-neutral. Conditioning on auction heterogeneity state $k$, bidder valuations are $i.i.d.$ draws from the same distribution $\Phi^k(v)$ with support $[\underline{v},\overline{v}]$. All bidders know auction heterogeneity $k$ before they submit bids. We denote the optimal bid distribution for state $k$ as $F^k(x)$, where $x$ is the optimal bid with support $[\underline{x},\overline{x}_k]$. For the remainder of this section and section 3, we present results that condition on a realization of the unobserved state. In Section 4, we then return to how we identify the state-specific bid distribution, i.e., the bid distribution conditioning on the UH, and the marginal distribution of UH.

    \subsection{The Ascending Auction Model} \label{OS2Parent}
    
Once the auction starts, bidders place their respective bids in ascending order. This process continues until there are no higher bids. Optimal bidding behaviors in these auctions are straightforward: a weakly dominant strategy is to continue bidding until the standing bid reaches your value.\footnote{Ascending auction in our content means ascending button auction, that is, we abstract from other possible behaviors, such as jump bidding in \citet{haile2003inference}.} In other words, once the next-to-last bidder drops out, the bidder with the highest value wins at a price equal to the second-highest value. We call the ``planned" dropout price the bid, which equals the value. Therefore, 
    \begin{eqnarray} \label{foc_2nd}
    \Phi^k(x) = F^k(x). 
    \end{eqnarray}

However, the highest bid is never observed. As a result, the distribution of the transaction price given auction state $k$, denoted as $F_{n-1:n}^k(\cdot)$, is the distribution of the $(n-1)^{th}$ order statistic out of an $n$ ordered sample. If the state $k$ and the number of bidders $n$ are known, the unknown state-specific bid distribution $F^k(\cdot)$ can be identified because there is a one-to-one mapping between the distribution of the $(n-1)^{th}$ order statistic and the underlying parent distribution itself.\footnote{That is, $F_{n-1:n}^k(x) = n(n-1)\int_0^{F^k(x)} t^{n-2} (1-t) dt.$} 

%Otherwise, we recover the state-specific bid distribution from the joint distributions of the observed dropout prices, and then we identify the state-specific value distribution. 

    \subsection{The First-Price Auction Model}
    
    In a first-price auction, the bidder with the highest bid wins and pays his own bid price. In a state-$k$ auction, a bidder with valuation $v$ chooses his bid $x$ to maximize his expected payoff, as follows: 
    \[
    \max_x \quad (v-x) \cdot \underbrace{\Phi^k \big(\xi^k(x) \big)^{n-1}}_{\text{probability of winning}}, 
    \]
    where $\xi^k(\cdot)$ is the inverse of his optimal bidding strategy in state-$k$ auctions, and $\Phi^k \big(\xi^k(x) \big)^{n-1}$ is the probability of winning. 
    
    \citet{GPV2000} studies the identification of the state-specific value distribution $\Phi^k(\cdot)$ when the competition level $n$ and the state-specific bid distribution $F^k(\cdot)$ are known. In particular, the identification comes from a one-to-one mapping between the unknown state-specific value distribution and the observed state-specific bid distribution: 
    \begin{equation}  \label{foc0}
    \xi^k(x) = x + \frac{1}{n-1} \frac{F^k(x)}{f^k(x)}, 
    \end{equation}
    where $x$ is any arbitrary bid in its support, and $F^k(\cdot)$ and $f^k(\cdot)$ are the state-specific bid distribution and density function, respectively. Once the state-specific bid distribution $f^k(\cdot)$ and competition $n$ is known, we can recover values corresponding to bids in the data and, in this way, identify the value distribution.

\paragraph{Remark} In both models, we need two elements to identify the state-specific value distributions $\Phi^k(\cdot)$: state-specific bid distributions $F^k(\cdot)$ and number of bidders $n$. In this paper, we will provide identification results for each case of when the number of bidders is known and unknown (to the analyst). For purposes of exposition, we start with assuming known competition and then extend the result to unknown competition. Since \eqref{foc_2nd} and \eqref{foc0} provide explicit mapping from the bid distribution to the value distribution, we can claim identification as soon as we identify state-specific bid distribution $F^k(\cdot)$ and number of bidders $n$. 
        
    \section{Order Statistics} \label{OS}
    
    We now derive a useful property of order statistics that we exploit to obtain our identification results. In particular, we show that the joint distribution of consecutive order statistics has a semi-separable structure with a state-independent indicator function that captures the correlation. We omit UH in this section.
    
    We first introduce some notation. Let $X_{1:n} \le X_{2:n}\le...\le X_{n:n}$ represent the $n$ order statistics out of an $n$ ordered sample. Let $f_{r,s:n}(\cdot,\cdot)$ denote the joint PDF of order statistics $\{X_{r:n}, X_{s:n}\}$; and let $f_{r:n}(\cdot)$ denote the marginal distribution of order statistics $X_{r:n}$. CDFs are defined analogously. 
    
    \subsection{Separability by Consecutiveness} \label{consecutive}
    
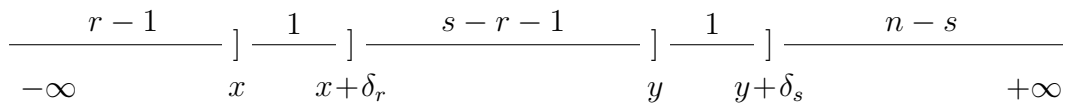
\begin{figure}[htbp]
    \centering
    \caption{Joint Distribution of Two Consecutive Order Statistics} \label{jdOS}
    \begin{tikzpicture}
    
    \draw 
    (0,0) -- node [above] {$r-1$} ++(3,0) node (rightRB1) [fill=white]{$]$}
    -- node [above] {$1$} ++(1.5,0) node (rightRB2) [fill=white]{$]$}
    -- node [above] {$s-r-1$} ++(4,0) node (rightRB3) [fill=white]{$]$}
    -- node [above] {$1$} ++(1.5,0) node (rightRB4) [fill=white]{$]$}
    -- node [above] {$n-s$} ++(4,0) node (fenhao) [fill=white]{} coordinate (rightend);
    
    \node [below] (2) at(rightRB1.south) {$x$};
    
    \foreach \i/\j in {rightRB2/x\!+\!\delta_r,rightRB3/y,rightRB4/y\!+\!\delta_s}
    \node [anchor=base] at(\i |- 2.base) {$\j$};
    
    \node [right] at(0,0 |- 2) {$-\infty$};
    \node [left] at(rightend |- 2) {$+\infty$};
    
    \end{tikzpicture}
    \end{figure}

    Suppose that, out of a sample of $n$ bids in the same auction, we observe two order statistics, $X_{r:n}$ and $X_{s:n}$, where $r<s$.  First, following \citet{david2004order}, we visualize the event $(x<X_{r:n} \leq x+\delta_r, y< X_{s:n} \leq y+\delta_s)$ with $x < x+\delta_r \le y$ in Figure \ref{jdOS}. When $\delta_r$ and $\delta_s$ are both small, the likelihood of this event is approximately proportional to the product of the following components: (1) the probability that $r-1$ draws are less than $x$, i.e., $[F(x)]^{r-1}$; (2) the probability that one draw is in $(x, x+\delta_r)$, i.e., $f(x) \delta_r$; (3) the probability that $n-s$ draws are greater than $y$, i.e., $[1-F(y)]^{n-s}$; (4) the probability that one draw is in $(y, y+\delta_s)$, i.e., $f(y) \delta_s$; (5) the probability that $s-r-1$ draws are in $(x + \delta_r, y)$, i.e., $[F(y) - F(x)]^{s-r-1}$. Second, we observe that the likelihood of this event can be represented in a semi-separable structure with respect to $x$ and $y$ if and only if $s-r-1=0$; i.e., the two order statistics are consecutive.
      
    \begin{lemma} \label{cOS}
Consecutive order statistics $X_{r-1:n}$ and $X_{r:n}$ have a semi-separable joint density with an indicator function capturing the correlation
\begin{eqnarray}
\label{eq0}
f_{r-1,r:n}(x,y) =c_{r-1,n} \cdot f_{r-1:r-1}(x) \cdot f_{1:n-r+1}(y) \cdot I(x\leq y) ,
\end{eqnarray}
where $c_{r-1,n} = \frac{n!}{(r-1)!(n-r+1)!} $ and $I(\cdot)$ represent the indicator function.
    \end{lemma}

This lemma formalizes the ``separability by consecutiveness'' shown in Figure \ref{jdOS}. $f_{r-1:r-1}(x)$ is the PDF of the top-order statistic of a sample of size $r-1$, and $f_{1:n-r+1}(y)$ is the PDF of the bottom-order statistic of a sample of size $n-r+1$. Compared to the joint PDF of (any) two random variables, a simple indicator function captures the correlation among two consecutive order statistics. As a result, their joint PDF is separable on the segment of $x \le y$. 

\paragraph{Remark} It is worth noting that the multiplicatively separable structure of the joint distribution of two consecutive order statistics, represented in Equation \eqref{eq0}, is different from the conditional independence of unordered bids. First, the joint distribution of any two unordered bids can be represented as the product of the marginal distribution of the two unordered bids. In contrast, the separable structure provided by the consecutiveness of the order statistics is represented by the product of the marginal distribution of two newly constructed random variables instead of the marginal distribution of the two ordered statistics. That is, $f_{r-1,r:n}(x,y)  \neq c_{r-1,n} f_{r-1:n}(x) f_{r:n}(y).$ Second, the multiplicatively separable structure only holds when $x\leq y$ and not at all times. If $x >y$, the left-hand side equals 0 by definition while the two marginal distributions do not necessarily equal 0. Therefore, we need to use the indicator function to capture the ordered relationship. However, we show that conditional independence is not necessary for identification with UH. Instead, the correlation  structure by the consecutiveness of the order statistics is sufficient for identification. 
   
\subsection{Distributions of Order Statistics}

    In general, the distribution function of the $r^{th}$ order statistic $X_{r:n}$ is 
    \[
    F_{r:n}(x) = \frac{n!}{(n-r)! (r-1)!} \int_0^{F(x)} t^{r-1} (1-t)^{n-r} dt , 
    \]
    which is strictly increasing in $F(x)$ and thus invertible. Therefore, there exists a one-to-one mapping between the distribution of any order statistic and its parent distribution $F(x)$. That is to say, the distribution of any order statistic is sufficient to identify its parent distribution. This property has been used in several papers including \citet{AH2002} and \citet{S2004}. Of course, this relationship between the distribution of an order statistic and its parent distribution depends on $n$, which is unobserved if we consider unknown competition.

\subsection{Auctions Versus Procurements} \label{OSauc}

As mentioned above, auction data often only capture a subset of all bids. In sealed-bid first-price auctions, we typically observe the most competitive bids, be it the highest bids in regular auctions or the lowest bids in procurement auctions. Note that we define the order statistics as $X_{1:n} \le X_{2:n}\le...\le X_{n:n}$, indicating that there are some subtle differences in notation when it comes to auctions and procurements. Specifically, using the language of order statistics, the winning bid is $X_{n:n}$ in regular auctions, i.e., $r=n$, but $X_{1:n}$ in procurement auctions, i.e., $r=1$.

These subtleties are more evident when the number of bidders is unknown, i.e., $n$ is unobserved. Specifically,  the two most competitive bids should be denoted as $\{X_{1:n},X_{2:n}\}$ in a procurement auction but as $\{X_{n:n}, X_{n-1:n}\}$ in a regular auction. As a result, we know the exact value of $r$ in the former, i.e., $r=1,2$, but only the relation between $r$ and $n$ in the latter, i.e., $r=n, n-1$. 

%Fortunately, there is a simple duality between the two scenarios: for an $n$ ordered sample of a random variable $X$, knowing the top two order statistics is equivalent to knowing the bottom two order statistics of the negation $-X$. 

%Consequently, when the number of bidders is unknown, we present results using the order statistics of the bids in procurement auctions or bid negations in regular auctions. 
  
\section{Main Results} \label{mainid}
    
    In this section, we consider identification of symmetric auction models with finite nonseparable UH using only order statistics of bids instead of all bids. First, conditional on the level of competition, we show that the cardinality of UH's support is identified. We then provide sufficient conditions to identify the state-specific value distributions using two consecutive order statistics and one instrumental variable. This result can be extended to the case with an additional order statistic but no instrumental variable. Second, we extend the identification result to allow for unobserved competition for ascending auctions. Last, we discuss estimation and inference. 

Suppose that, for a given auction, the bidder values are i.i.d.\ draws from the same distribution $\Phi^k(v)$, where auction state $k$ is finite and discrete, i.e., $k=1,...,K$, and state $k$ is realized with probability $p_k>0$, and $\sum^K_{k=1}  p_k=1$. With slight abuse of notation, we will consistently use $c$ to denote a known constant and $\eta$ to denote an unknown constant throughout the paper. The econometrician observes two consecutive bids  and one instrumental variable, i.e., $\{(X_{r-1:n}, X_{r:n}),W\}$, while the auction-level state $k$ is unobserved. Therefore, any component that only involves the order statistics of bids are directly estimable from the data and thus can be treated as known while any state-specific component is to be identified.
 \label{woFOSD}
    
\subsection{Identification with Known Competition}    

%This subsection uses the abovementioned property of consecutive order statistics to identify auction models with a finite and discrete unobserved state when the competition is known. That is, the identification results are conditional on the competition. The identification mainly exploits the semi-separable structure of the joint distribution of consecutive order statistics. We first show that the cardinality of the support of UH can be identified using the rank of the observed matrix constructed by the joint distribution of any two consecutive order statistics. We then show that with an additional instrument, even being a binary one, we can identify the uh-specific bid distribution and thus the associated state-specific value distribution. \textcolor{red}{[RX: is this paragraph repetitive of the paragraph right after Section 4?]}

    The unconditional joint PDF of two consecutive order statistics can be estimated directly from the data. By total probability, it can be represented as
    \begin{eqnarray}
f_{r-1,r:n}(x,y)
    &=& \sum^K_{k=1}  p_k f^k_{r-1,r:n}(x,y) \notag\\
  &=& \sum^K_{k=1}  p_k c_{r-1,n}  f^k_{r-1:r-1}(x) f^k_{1:n-r+1}(y) \cdot  I(x \le y),      \label{eq_joint1} 
    \end{eqnarray}
where superscript $k$ indicates the state-$k$ parent distribution. Note that $c_{r-1,n} = \frac{n!}{(r-1)!(n-r+1)!} $,  $f_{r-1,r:n}(x,y)$ can be directly estimated from the data while $p_k$, $f^k_{r-1:r-1}(x)$, and $f^k_{1:n-r+1}(y)$ are the unobserved component to be identified. This indicator function, capturing the correlation of the order statistics, precludes us from directly following the existing procedure of eigenvalue-eigenvector decompositions to identify the state-specific bid distribution. Fortunately, such a correlation is known and does not depend on the unobserved state, which enables us to modify the existing identification result to achieve identification in our context. 

Specifically, if we limit variation in the order statistics in predetermined no-overlapping intervals, the ordering correlation trivially holds. Essentially, to account for this correlation/ordering, we introduce a discretization of bids before proceeding to identification analysis. We divide the original support into two segments, ``low" and ``high," using one cutoff $\xL$, where $\underline x<\xL<\bar x = \max_k \{\bar x_k\}$. Denote the two segments as $l = [\underline{x}, \xL]$ and $h = [\xL, \bar x]$. Therefore, if we only exploit the joint variation of $X_{r-1:n}=x$ and $X_{r:n}=y$ in the two ordered interval $x\in l$ and $y\in h$, the separable structure of the joint PDF $f_{r-1,r:n}(x,y)$ reappears, which has three important features. First, this representation has a similar structure to finite mixture models. This feature allows us to turn this structure into an eigendecomposition to study identification. Second, the variations of the two observed ordered bids are restricted to their respective intervals. This restriction requires us to study the identification of component distributions interval by interval. Third, each term in the multiplicatively separable structure is the PDF of a newly constructed order statistic associated with the parent distribution. This allows us to translate the identified distributions of order statistics into the parent distributions. 

%    \begin{eqnarray}     f_{r-1,r:n}(y,z) \!= \sum^K_{k=1}  p_k f^k_{r-1:r-1}(y)  \cdot c_{r,n} f^k_{1:n-r+1}(z),     \label{eq_joint1} \end{eqnarray}

\begin{figure}[h!]
    \centering
    \caption{Discretization} \label{disfig}
    \begin{tikzpicture}[scale=0.750]
    
    \draw [thick] (0,0) -- node[above]{{\footnotesize$l_1$}} ++(1.5,0) coordinate (tick1)
    -- node[above]{{\footnotesize$l_2$}} ++(1.5,0) coordinate (tick2)
    -- ++(3,0) coordinate (tick3)
    -- node[above]{{\footnotesize$l_{\tilde K}$}} ++(1.5,0) coordinate (tick4)
%    -- node[above]{{\footnotesize$m_1$}} ++(1.5,0) coordinate (tick5)
%    -- node[above]{{\footnotesize$m_2$}} ++(1.5,0) coordinate (tick6)
    -- node[above]{{\footnotesize$h_1$}} ++(1.5,0) coordinate (tick5)
    -- node[above]{{\footnotesize$h_2$}} ++(1.5,0) coordinate (tick6)
    -- ++(3,0) coordinate (tick7)
    -- node[above]{{\footnotesize$h_{\tilde K}$}}++(1.5,0) coordinate (rightend);

    \foreach \i in {1,...,7}
    \draw (tick\i)--+(0,1mm);

    \draw (0,0)--+(0,5mm);
    \draw (7.5,0)--+(0,5mm);
   % \draw (10.5,0)--+(0,5mm);
    \draw (rightend)--+(0,5mm);

    \pgfkeys{/tikz/bracetemp/.code n args={4}{
    \tikzmath{
    coordinate \adjvset,\bracesp;
    \adjvset=#4;
    \bracesp1=#1;
    \bracesp7=#2;
    \bracesp4=($(\bracesp1)!0.5!(\bracesp7)+#3$);
    \bracesp2=(\bracesp1)+(\adjvset);
    \bracesp3=(\bracesp4)-(\adjvset);
    \bracesp5=(\bracesp4)+(\adjvsetx,-\adjvsety);
    \bracesp6=(\bracesp7)-(\adjvsetx,-\adjvsety);
    }
    }}

    \draw [decorate,blue, decoration={brace,amplitude=10pt,mirror}](0,0) -- (7.5,0) node [black,midway,yshift=-0.65cm]{\footnotesize $x \in l $};
%    \draw [decorate,blue, decoration={brace,amplitude=10pt,mirror}](7.5,0) -- (10.5,0) node [black,midway,yshift=-0.65cm]{\footnotesize $y \in m$};
    \draw [decorate,blue, decoration={brace,amplitude=10pt,mirror}](7.5,0) -- (15,0) node [black,midway,yshift=-0.65cm]{\footnotesize $y \in h$};

    \filldraw 
    (0,0) circle (2pt) node[align=left,   below] {$\underline x$} --
    (7.5,0) circle (2pt) node[align=center, below] {$\xL$}     -- 
%    (10.5,0) circle (2pt) node[align=right,  below] {$\xR$} --
    (15,0) circle (2pt) node[align=right,  below] {$\overline x$};
    
    %\draw [bracetemp={(0,0)}{(tick4)}{(0,-0.4)}{(0.1,-0.4)}](\bracesp1) .. controls (\bracesp2) and (\bracesp3) .. (\bracesp4) node(xr-2)[below]{$x$} .. controls (\bracesp5) and (\bracesp6) .. (\bracesp7);
    
    %\draw [bracetemp={(tick4)}{(tick6)}{(0,-0.4)}{(0.4,-0.4)}](\bracesp1) .. controls (\bracesp2) and (\bracesp3) .. (\bracesp4) node(xr-1)[below]{$x_{r-1}$} .. controls (\bracesp5) and (\bracesp6) .. (\bracesp7);
    
    %\draw [bracetemp={(tick6)}{(rightend)}{(0,-0.4)}{(0.1,-0.4)}](\bracesp1) .. controls (\bracesp2) and (\bracesp3) .. (\bracesp4) node(xr)[below]{$x_{r}$} .. controls (\bracesp5) and (\bracesp6) .. (\bracesp7);
    
    %\foreach \i/\j in {{0,0}/\underline{x},{tick4}/\underline{b_2},{tick6}/b_2,{rightend}/\overline{x}}
    %\node [anchor=base] at (\i |- xr.base) {$\j$};
    
    \end{tikzpicture}
    \end{figure}
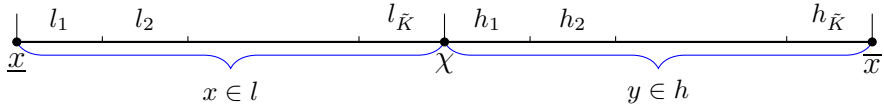

Similar to the finite mixture and measurement error literature (\citet{H2008}), our identification uses matrix algebra. We now further discretize the bid support and introduce some matrix notation. We select $\tilde K$ exclusive intervals from ``low" segment $l$ and ``high" segment $h$, denoted as $l_i$ and $h_i$, respectively, where $i=1,...,\tilde K$. Note that these intervals do not have to be fully exhaustive, and they can simply be points. Figure \ref{disfig} provides a visualization of the discretization. 
    
 Following Equation (\ref{eq_joint1}), we express the probability of the event $\{X_{r-1:n} \in l_i,X_{r:n} \in h_j\}$ as 
    \begin{eqnarray*}
   \Pr(X_{r-1:n} \in l_i, X_{r:n} \in h_j)      &=&  \sum^K_{k=1}  \int_{x \in l_i}  \int_{y \in h_j }p_k  f^k_{r-1:r-1}(x) c_{r-1,n}  f^k_{1:n-r+1}(y) dx dy   \notag \\
    &=&   \sum^K_{k=1}    \int_{x \in l_i}  f^k_{r-1:r-1}(x)d x  \cdot p_k  \cdot \int_{y \in h_j } c_{r-1,n}  f^k_{1:n-r+1}(y) d y.
    \end{eqnarray*}
We first introduce the following matrix notation: 
    \begin{eqnarray*}
    \boldsymbol{J}_{l,h}&=& \{ \Pr(X_{r-1:n} \in l_i , X_{r:n} \in h_j) \}_{i,j}, \textrm{for } i,j=1,...,\tilde K, \\
    \boldsymbol{L} &=&  \{ \int_{x \in l_i }     f^k_{r-1:r-1}(x)d x \}_{i,k}, \textrm{for } i,k=1,...,\tilde K,\\
%    \boldsymbol{D}_{m_{i'}} &= & diag \{\int_{y \in m_{i'}}f^k(y) d y\}_k , k=1,..., K \\
    \boldsymbol{D}_p &= & diag \{p_k\}_k, \textrm{for } k=1,..., \tilde K,\\
    \boldsymbol{H} &=& \{ \int_{y \in h_j}c_{r-1,n} f^k_{1:n-r+1}(y) d y \}_{j,k}, \textrm{for } j,k=1,...,\tilde K,
    \end{eqnarray*}
    where $\boldsymbol{J}_{l,h}$ is a probability matrix with $(i,j)^{th}$ element representing the probability of event $\{X_{r-1:n} \in l_i,X_{r:n}\in h_j\}$, which can be identified and estimated directly from the data. $\boldsymbol{L}$ is a probability matrix with $(i,k)^{th}$ element representing the probability of event $\{X^k_{r-1:r-1} \in l_i\}$;  $\boldsymbol{D}_p$ is a diagonal matrix with $k^{th}$ diagonal element representing the weight of state $k$; $\boldsymbol{H} $ is a probability matrix with $(j,k)^{th}$ element representing the probability of event $\{X^k_{1:n-r+1} \in h_j\}$. Note that superscript $k$ indicates the random variable is associated with the state-$k$ parent distribution. Also, matrices $\boldsymbol{L}$, $\boldsymbol{D}_p$, and $ \boldsymbol{H}$ are unobserved and yet to be identified. 
    
    With the matrix notation, we have the following matrix representation connecting the observed joint probability with the unknown component matrices:
    \begin{eqnarray}
    \label{eq_joint_M1}
    \boldsymbol{J}_{l,h}= \boldsymbol{L} \boldsymbol{D}_p \boldsymbol{H}^T. 
    \end{eqnarray}
    
    Following the existing literature (\citet{H2008}), identification of models with UH using the mixture features generally requires a rank condition, which is stated as follows.
\begin{customasp}{FR} (Full Rank)
    There exists a cutoff of the support so that the distributions of $f^k_{r-1:r-1}(x)$, for $k=1,...,K$, where $x \in [\underline x, \xL]$, are linearly independent, and the distributions of $f^k_{1:n-r+1}(y)$, where $y\in [\xL, \bar x]$ and $k=1,...,K$, are linearly independent.\footnote{Note that the full rank condition requires that the intersection of the supports of different states is nonempty. In our context, this trivially holds since all states have the same lower bound. }
    \label{ass_fullrank1}
\end{customasp}
The linearly independent assumption basically requires that there is sufficient variation in the bid distribution for different unobserved states. The smaller $K$ is, the easier this condition holds. For example, if $K=2$, this assumption trivially holds for any two bid distributions except that both state-specific bid distributions are uniform distributions.

\paragraph{Identification of $K$} We first show that we can identify the unobserved cardinality of the support of UH; that is, $K$ is identified. In contrast, the existing literature usually assumes $K$ is known, as in \cite{HMS2013}. We explicitly use superscript $\tilde K$ to denote  the matrices associated with a $\tilde K$-interval discretization of $l$ and $h$. That is, the joint distribution of the two order statistics with $\tilde K$-interval discretization can be represented as
    \begin{eqnarray}
    \boldsymbol{J}^{\tilde K}_{l,h}= \boldsymbol{L}^{\tilde K} \boldsymbol{D}_p {\boldsymbol{H}^{\tilde K}}^T,
    \end{eqnarray}
where the observed matrix $\boldsymbol{J}^{\tilde K}_{l,h}$ has dimensions of $\tilde K \times \tilde K$, the matrices $ \boldsymbol{L}^{\tilde K} $ and $\boldsymbol{H}^{\tilde K}$ have dimensions $ \tilde K \times K$, and $\boldsymbol{D}_p$ are diagonal matrices with dimension of $K \times K$. 

We now derive the relationship between the unknown $K$ and the rank of  the joint probability matrix $ \boldsymbol{J}^{\tilde K}_{l,h}$, which is directly estimable from the data. 

\begin{lemma}\label{id_K} If Assumption \ref{ass_fullrank1} holds, the maximum of $rank(\boldsymbol{J}^{\tilde K}_{l,h})$ among all $\tilde K$-interval discretization equals $\min \{\tilde K, K\}$. 
\end{lemma}

%When $\tilde K \le K$, the maximum of $rank(\boldsymbol{J}^{\tilde K}_{l,h})$ among all $\tilde K$-interval discretization equals $\tilde K$, so it is strictly increasing in $\tilde K$; and when $\tilde K \geq K$, it equals $K$. 

Lemma \ref{id_K} says that the maximum of $rank(\boldsymbol{J}^{\tilde K}_{l,h})$ among all $\tilde K$-interval discretization is strictly increasing in $\tilde K$ when $\tilde K < K$ and constant when $\tilde K \geq K$. 
Intuitively, one can start with $\tilde K=2$ and stops until the rank of the observed matrix $\boldsymbol{J}^{\tilde K}_{l,h}$ stops growing. Once the unobserved $K$ is identified, we treat it as known and suppress the discretization  of $K$ as superscript hereafter.

\paragraph{Identification of the State-specific Bid Distribution}Once the cardinality of UH, i.e., $K$, is identified, we show that the state-specific bid distribution can also be nonparametrically identified if there exists an instrumental variable, which we denote as $W$. The requirement for the instrument is mild --- as long as there is variation in the instrument; that is, the instrumental variable can be a binary variable.\footnote{This is similar to the \citet{hu2017econometrics} 2.1 measurement model.} We present the assumptions for a valid binary instrumental variable. All assumptions and identification results here can be readily extended to the situation with general instrumental variables.
\begin{customasp}{IV}A valid instrumental variable must satisfy the following two conditions.\\
1). (exogeneity)  Conditional on UH, the instrument reveals no information about the bids and the associated order statistics.  \\ 
2). (relevance) The unobserved state reveals distinct information about the instrumental variable $W$.\footnote{When $W$ is not a binary variable, an alternative  relevant condition is that $E(W|k) \neq E(W|k'), \forall ~ k \neq k',$ which indicates that the UH reveals information about the mean of the instrumental variable.} That is, $  \Pr(W=0|k) \neq  \Pr(W=0|k'),~ \forall ~ k \neq k'.$
    \label{ass_distinct}
%    \begin{eqnarray*}
%    \frac{\int_{y \in m_{1}}f^k(y) dy}{\int_{y \in m_{2}}f^k(y) d y} \neq \frac{\int_{y\in m_{1}}f^{k'}(y) d y}{\int_{y \in m_{2}}f^{k'}(y) d y}, \quad \forall \quad k \neq k'. 
%    \end{eqnarray*}
\end{customasp}
In empirical applications, we can often construct such an instrument from supplementary data sources. For example, one possible instrumental variable in timber auctions is slope (i.e., the incline of the land) or aspect (i.e., the compass direction that a terrain surface faces). Because the department of natural resources geocodes the locations of timber lots, we can derive both slope and aspect using GIS and public elevation data. For example, in the Northern Hemisphere, timber on the southern side receives more sunlight than on the northern side. As a result, timber on the southern side tends to be of higher quality $k$, i.e., satisfying the relevance condition. On the other hand, the aspect does not directly affect bidders’ valuations, satisfying the exogeneity condition. Similar examples include average rainfall or soil quality (\cite{HMS2013}). In highway procurement auctions, projects in areas with high traffic volume tend to be of higher difficulty. On the other hand, traffic volume does not directly affect bidders' costs.  

With the instrumental variable's exogeneity condition being satisfied, the joint distribution of the two consecutive order statistics, i.e., $X_{r-1:n}=x$ and $X_{r:n}=y$ in the two ordered interval $x\in l$ and $y\in h$, and the instrumental variable can be represented as 
    \begin{eqnarray}
    f_{r-1,r:n}(x,y, W) &=&  \sum_k p_k f^k_{r-1:r-1}(x) \cdot c_{r-1,n}  \Pr(W|k) \cdot f^k_{1:n-r+1}(y).
%    f_{r-1,r:n}(y,z) &=&\frac{n! I(y \le z)}{(r-1)!(n-r+1)!}  \sum_k p_kf^k_{r-1:r-1}(y) \cdot f^k_{1:n-r+1}(z) .
    \end{eqnarray}
Fixing $W=0$, using the $K$ intervals in the $l$ and $h$ segments, we can rewrite the above equations connecting the unknown state distribution with the observed joint probability into a matrix representation:
    \begin{eqnarray}
    \label{eq_joint_M10}
    \boldsymbol{J}_{l,h,0}= \boldsymbol{L} \boldsymbol{D}_p \boldsymbol{D}_0 \boldsymbol{H}^T,
    \end{eqnarray}
where $\boldsymbol{J}_{l,h,0}$ is constructed similar to $\boldsymbol{J}_{l,h}$ with $W=0$, and $\boldsymbol{D}_0$ is a diagonal matrix with the $l^{th}$ diagonal element being  $\Pr(W=0|k)$.

If  Assumption \ref{ass_fullrank1} holds, there exists a discretization so that matrices  $\boldsymbol{J}_{l,h}$ are invertible because both $\boldsymbol{L}$ and $\boldsymbol{H}$ are full rank. Therefore, combinig Equations \eqref{eq_joint_M1} and \eqref{eq_joint_M10} leads to the following main equation for identification:
    \begin{eqnarray}
    \label{main_eq_3order}
    \boldsymbol{J}_{l,h,0} \boldsymbol{J}^{-1}_{l,h}= \boldsymbol{L}   \boldsymbol{D}_{0} \boldsymbol{L}^{-1} ,
    \end{eqnarray}
which indicates that the observed matrix on the left-hand side and the unknown matrices on the right-hand side are similar. Specifically, probability matrix $ \boldsymbol{L}$ of the ``low" segment and the conditional probability $ \boldsymbol{D}_{0} $ can be identified as the eigenvector and eigenvalue matrices of the observed joint probability matrix, respectively. 

The relevance of the instrumental variable guarantees that the decomposition admits distinct eigenvalues, which guarantees  the eigendecompositon to be unique. The identification of the state-specific bid distribution then proceeds in several steps. First, an eigenvalue decomposition argument identifies a key matrix that governs the finite mixture structure in our order statistic setting. This allows for the identification of component bid distributions using the one-to-one mapping between the distribution of an order statistic and its parent distribution in the low and high segment of the support. Second, we pin down the scales. 

%We summarize these identification results in the following lemmas. 
    
    \begin{lemma} [Eigendecomposition] \label{id_main}
    If Assumptions (\ref{ass_fullrank1}) and (\ref{ass_distinct}) are satisfied, the state-specific distribution in the ``low" and ``high" segments, i.e., $\boldsymbol{L}$ and $\boldsymbol{H}$, are identified up to the same permutation of the columns but different scales.\footnote{That is, we identify $\boldsymbol{L} Q \Lambda_\ell $ and $ \boldsymbol{H} Q \Lambda_h$, where $\Lambda_\ell$ and $\Lambda_h$ are two diagonal matrices and $Q$ is an elementary matrix generated by interchanging columns of the identity matrix.} Moreover, the probability of the instrument being 0 conditional on the UH is also identified up to the same permutation.
    \end{lemma}
Identification of the components up to permutation is a prevalent feature of identification via eigendecomposition, which requires additional information about the UH to pin down its exact value. If one is unwilling to impose such an assumption, we can keep the UH as an index without providing any economic meaning to the labeling. Continuing the example of aspect, suppose it maps to UH one-to-one. Since timber sales are a mixture of the two unobserved states, Lemma \ref{id_main} means that we can identify two sets of objects --- each with the probabilities associated with the same state in ``low'' and ``high'' segments --- but we cannot determine which set is for the southern or northern side. That is, the ordering of the UH is the same in both matrices. Typically, a monotonicity condition can pin down the true labeling of the UH, which depends on the economic context of the UH. Knowing that timber on the southern side tends to be more valuable allows us to assign aspect to the identified objects.

%we can identify $K$ component distributions but are agnostic about which component distribution associated with which component $k$. Moreover, $\boldsymbol{H}$ can be identified up to the same permutation (with $\boldsymbol{L}$). For example, in timber auctions aspect can be a critical UH: in the northern hemisphere, the timber on the southern side receives more sunlight than the north side. If timber sales are a mixture of the two unobserved types, Lemma \ref{id_main} says that we can identify two sets of objects, knowing that each of them is from the same side, but we could not tell which side generates which distributions.Typically a monotonicity condition is required to pin down the true labeling of the UH, which depends on UH's economic context. 

Identification up to scales means that the component distribution identified from the decomposition is not the distribution itself but a scalar multiplication of it. This feature is also prevalent in identification using the mixture feature. Pinning down the scale requires a normalization condition. In the existing literature of independent measurements, such as \cite{HMS2013},  since the identification exploits the variation of each bid in its full support, such a feature enables direct pinning down the scales. Specifically, each column of their eigenvector matrix $\boldsymbol{L}$ sums to one because this sum represents the cumulative distribution over the full support. However, such a normalization condition is not applicable in our order statistics context. This is because, to guarantee the multiplicative separable structure, we can only exploit the variation of ordered bids in two exclusive segments of the full support. Therefore, each column of our eigenvector matrix $\boldsymbol{L}$ sums to an unknown quantity strictly less than one, representing the probability of observing $X_{r-1:r-1}^k$ within the ``low'' segment, i.e., $F^k_{r-1:r-1}(\xL)$.

To summarize, we identify the state-specific bid distributions in the ``low" and ``high" segments up to the same ordering but different scales: 
    \begin{eqnarray}
  f^{k}(x)=  \begin{cases}
    & \eta_l^k \check{f}_l^{k}(x) \quad  \hbox{if}\hspace{0.2cm} x \leq \xL, \\
 & \eta_h^k \check{f}_h^{k}(x) \quad  \hbox{if}\hspace{0.2cm} x \geq \xL,
    \end{cases}%
    \label{id_sum0}
    \end{eqnarray}  
where $\check{f}_l^{k}(\cdot)$ and $\check{f}^{k}_h(\cdot)$ are the components identified for the ``low"  and ``high" segments with the associated scales $\eta_l^k$ and $\eta_h^k$, respectively.  

To identify the state-specific value function, we need to pin down the scales, which requires two restrictions for the two unknowns for each $k$. We invoke the continuity of the component PDFs and the total probability argument. First, the PDFs identified separately in the ``low" and ``high" segments should be the same at cutoff point $\xL$ due to the continuity of the true component PDF. Second, the fact that each component PDF integrates to 1 provides the second restriction on the scales. In the Appendix, we show that these restrictions uniquely identify the scales. We summarize the result that scales are identified in the following lemma.
    
    \begin{lemma} (Scales)
    \label{lemma_scales}
    The state-specific scales associated with the state-specific distributions of the ``low"  and ``high" segments are identified.  
    \end{lemma}
    Once the scales have been pinned down, we can identify state-specific weight $p_k$ using the marginal PDF/CDF of one order statistic. To summarize, we can identify the state-specific bid distribution using only two consecutive order statistics of the bids and one binary instrumental variable with suitable rank conditions. After identifying the state-specific bid distribution, we then identify the state-specific value function for first-price and ascending IPV auctions since the number of bidders is known. The following theorem summarizes our results. 
    
\begin{theorem}
\label{theorem1}
If competition $n$ is known and Assumptions (\ref{ass_fullrank1}) and (\ref{ass_distinct}) are satisfied, then state weight $p_k$, state-specific bid distribution $f^k(x)$ for $x \in [\underline x, \bar x_k]$, and state-specific value distribution $\Phi^k(v)$ for $v\in[\underline v, \bar v]$ are identified up to permutation for all unobserved states using two consecutive order statistics with a binary instrumental variable under both first-price and ascending IPV auctions.
\end{theorem}

\paragraph{Remark}Lemma \ref{cOS} implies that $X_{r-1:n}$ and $X_{r:n}$ are independent conditioning on the event $\{X_{r-1:n} \in l, X_{r:n} \in h\}$. Therefore, we can also consider $\Pr(X_{r-1:n} \in l_i, X_{r:n} \in h_j | X_{r-1:n} \in l, X_{r:n} \in h)$ and follow \citet{H2008}, as in \citet{HMS2013}, in conducting eigendecomposition to identify the two probability matrices and then the conditional distribution of the associated order statistics. \citet{HMS2013} does not require further treatment for scales. In contrast, there are two unresolved issues in our context. First, the decompositions can only identify the conditional distribution of the order statistics, and one needs to invoke the one-to-one mapping between the distribution of an order statistic and its parent distribution. Second, while the decompositions identify the conditional distribution of the order statistics, the probability of the conditional event is unknown. Pinning down the latter requires additional treatment, such as a strategy similar to Lemma \ref{lemma_scales}. 

It is also worth noting that while the literature mainly relies on the independence property for identification (\citet{H2008}), the eigendecomposition approach essentially exploits the implication of such a property --- multiplicative separability. That is, multiplicative separability is a condition weaker than independence that is sufficient for identification. This observation makes our finding ``separability by consecutiveness'' essential in solving the long-standing identification problem. Therefore, we work directly with the separability of the unconditional joint distribution of the consecutive order statistics, obviating the additional conditioning in the proofs.

\paragraph{Remark} When it is difficult to find a valid instrumental variable, we can use an additional order statistic of bids instead. Specifically, a similar logic can be applied to three consecutive order statistics $\{X_{r-2:n}, X_{r-1:n},X_{r:n}\}$; their joint PDF has the following semi-separable structure: 
    \begin{eqnarray}  \notag
f_{r-2, r-1,r:n}(x,y,z)=& \frac{n!}{(r-2)!(n-r+1)!} \underbrace{f_{r-2:r-2}(x) \cdot f(y)  \cdot f_{1:n-r+1}(z)}_{\text{multiplicatively separable}}  \cdot \underbrace{I(x\le y\le z )}_{\text{correlation}}.
    \label{eq_semi}
    \end{eqnarray}
Intuitively, we can treat the additional order statistic as the instrumental variable. See our companion paper \citet{luo2020order} for a formal discussion.
 
\subsection{Identification with Unknown Competition} \label{id_sec}

In this subsection, we consider two sources of unobserved auction-level characteristics: unobserved auction state $k$ and unobserved competition $n$. For simplicity, we focus on ascending auctions in this subsection.\footnote{See our working paper \citet{luo2020identification} for discussion on first-price auctions with unknown competition.} We build on \citet{S2004}, which considers online English auctions with unobserved competition. Related to our result, \citet{FL2017} studies the problem in the classical setting (i.e., \textit{separable and continuous} UH) using reserve price and two order statistics of bids; without identifying the value distributions, \citet{coey2021scalable} focuses on the identification of the optimal reserve price using the top two bids for online auctions under a second-price-auction-like format with sequential arrival of bidders. Our approach uses two consecutive order statistics with an instrument, and achieves point identification under \textit{nonseparable and finite} UH.

We assume exogenous participation, that is, the value distribution only varies with the auction state but not the number of bidders, and the number of bidders takes values from a known set, i.e., $n \in \{\underline N, \underline N+1, ..., \bar N\}$, where $\underline N$ and $\bar N$ are known.\footnote{In principle, we can further exploit model restrictions to identify the support. For instance, \citet{GL2018} proposes a density discontinuity approach in first-price auctions using the fact that the upper boundary of the bid distribution is strictly increasing in the number of bidders. However, this approach fails in ascending auctions. An alternative is to obtain estimates from other sources. For instance, the seller may require at least two bidders. Therefore, $\underline{N} = 2$. If the data contain firm locations, the maximum number of bidders $\overline{N}$ can be estimated by the number of firms in the local market.} Moreover, $p_{k,n} \in (0,1), \sum_{k,n} p_{k,n} =1$. Denote the number of possible competition levels as $|n|$, so $|n|=\bar N-\underline N+1$. Under exogenous participation, the value distribution differs with auction state $k$ but is the same regardless of competition $n$. That is, there are $K$ different value distributions to be identified, i.e., $\Phi^k(v)$, where $k=1,...,K$. In ascending auctions, bidding their true values is a weakly dominant strategy regardless of the competition level. That is, the optimal bidding strategy is the same for any number of bidders. Consequently, the state-competition-specific bid distribution is independent of $n$, i.e., $F^{k,n}(\cdot) = \Phi^{k}(\cdot)$. 

As discussed in Subsection \ref{OSauc}, there are some notation subtleties when competition is unknown. Here, we represent the $n$ order statistics as $X_{1:n}\geq X_{2:n} \geq \cdots \geq X_{n:n}$ in regular auctions and $X_{1:n}\leq X_{2:n} \leq \cdots \leq X_{n:n}$ in procurement auctions, which condenses the unknowns in the subscript to one element. Additionally, we can translate these results to procurement auctions by replacing all mention of ``value'' with ``cost.'' 

%identification results in this subsection apply to the order statistics of the negatives of the bids. We denote the negative bid as $X$ and record the observed ones in ascending order, i.e., $\{X_{r-1:n}, X_{r:n}\}$, where $r$ is known and $n$ is unknown. This concentrates the unknowns in the subscript into one element.\footnote{Using the order statistics of the bids, the winning bid would be $X_{n:n}$, where both elements in the subscript are unknown.} Note that the lowest order statistic corresponds to the winning bid, i.e., the winning bid equals $-X_{r-1:n}$

To understand the source of identification, we first demonstrate the identification abstracted from unobserved states and then extend the identification to allow for both unobserved competition and  unobserved state. 
\paragraph{Unobserved Competition} When the only unobserved factor is the number of bidders, the joint distribution of order statistics $X_{r-1}$ and $X_r$ can be represented as
\begin{eqnarray}
f_{r\!-\! 1,r}(x,y) 
&=&f_{r\!-\! 1:r\!-\! 1}(x) 
\underbrace{\sum_n \left[p_{n}  c_{r-1,n}  f_{1:n\!-\! r+1}(y)\right]}_{\text{mixture of competition}} 
I( x \le y) , \label{eq_second_1}
\end{eqnarray}
which follows from Lemma \ref{cOS}. Consecutive order statistics have a much simpler joint distribution than arbitrary ones, as employed in \citet{S2004}. This joint distribution reveals that the unconditional joint distribution from the data only has a mixture of order statistics larger than the $r^{th}$ while the information of the $(r-1)^{th}$ order statistic is invariant to the level of competition. \citet{S2004} makes this important observation. This is because the order statistics observed in the data are from the bottom. Intuitively, if we fix the $r^{th}$ order statistic to be $\xL$, the joint density of the two order statistics will be equivalent to the marginal distribution of the top order statistics from an $r-1$ sample scaled by an unknown constant, i.e., $\bar f(y) \equiv \sum_n p_{n}   c_{r-1,n}  f_{1:n-r+1}(y) $. Consequently, we can identify the marginal distribution for such a top order statistic out of an $r-1$ sample up to an unknown scale; i.e., $f_{r-1:r-1}(x)$ is identified with an unknown constant for $x\le \xL \le \bar x$. Furthermore, we also identify its parent distribution up to a scale using its connection with the distribution of its order statistic. That is, 
\begin{eqnarray*}
F(x) = \eta \check{F}(x), \forall x \le \xL \le \bar x,
\end{eqnarray*}
where $\check {F}(x)$ is identified and $\eta$ is the unknown scale.

To fully identify the bid distribution, we need to tackle the following two issues: (1) we need to pin down the scale $\eta$; (2) we need to identify the bid distribution for $y \ge \xL$. \citet{S2004} solves both problems simultaneously by taking $\xL$ to the upper limit so that the whole distribution is identified, because scale $\eta$ can be pinned down using the fact that $F(\bar x)=1$. 

We propose an alternative approach that simultaneously identifies the scales and distribution of competition $p_n$. Specifically, we combine both Equation \eqref{eq_second_1} and the one-to-one mapping between the CDF of the observed $r^{th}$ order statistic and its parent distribution. In fact, the distribution of the observed $r^{th}$ order statistic is a mixture 
\[
F_r(x) = \sum_n  p_n \sum_{i=r}^{n} c_{i,n} [\eta \check{F}(x)]^{i}[1-\eta \check{F}(x)]^{n-i}, \forall x \le \xL,
\] 
whose left-hand side is known and right-hand side is linear in $p_n$ and polynomial in $\eta$. It allows us to construct a system of $|n|+1$ equations that provides identifying restrictions on both $p_n$s and $\eta$. We summarize this result in the following lemma. 

    \begin{lemma} 
    \label{lemma_onlyn}
   If only the competition is unobserved, the value distribution and the distribution of competition can be identified using two consecutive order statistics.
    \end{lemma}

The distribution of competition is of interest in many settings. Moreover, our approach avoids taking $z$ to the upper limit, which brings advantages in the setting with two sources of unobserved factors.   

\paragraph{Two Sources of Unobserved Factors} We now examine the identification of ascending auctions allowing for both unobserved state and unobserved competition. The identification requires additional variation such as a binary instrumental variable besides the two consecutive order statistics because of the additional unobserved auction state $k$. 

The identification exploits the following properties: (1) bids are independent across bidders in the same auction; (2) the joint density of consecutive order statistics admits a semi-separable structure; (3) we can lump the impact of unobserved competition into one component that only involves information about the $r^{th}$ order statistic due to the auction format. That is to say, we can treat the mixture with two sources of unobserved factors $(k,n)$ as a mixture with one source of unobserved state $k$ by lumping together all competition effects. 

We first identify the state-specific density for the ``low" segment and the component mixture of all competition in the ``high" segment up to scales using eigendecomposition. We then follow the intuition for the scenario without unobserved states to pin down the scales and weight distribution and to identify the state-specific bid/value distributions. 

Specifically, the joint distribution of these two consecutive order statistics together with the instrumental variable can be represented as a mixture of state $k$ and competition $n$. Specifically, if we only exploit the variation of the two order statistics $X_{r-1:n}=x$ and $X_{r:n}=y$ in the two ordered intervals $x\in l$ and $y\in h$, we have
\begin{eqnarray}
f_{r-1,r}(x,y,W) 
%&=& \sum_{k,n} p_{k,n}   f^{k,n}_{r-2,r-1,r:n}(x,y,z)\notag\\
%&=&  \sum_k p_k \frac{n!}{(r-3)! (n-r)!} [F^k(x_{r-2})]^{r-3} f^k(x_{r-2}) f^k(x_{r-1})[1-F^k(x_r)]^{n-r}f^k(x_r)
%&=& \sum_{k,n} p_{k,n}  c_{r,n} f^{k}_{r-2:r-2}(x)  f^{k}(y) f^{k}_{1:n-r+1}(z) I(x \le y \le z)\notag\\
&=& \sum_{k}f^{k}_{r-1:r-1}(x) \cdot Pr(W|k) \cdot \underbrace{\sum_n p_{k,n}  c_{r-1,n}  f^{k}_{1:n-r+1}(y)}_{\text{mixture of competition}}.
\label{2nd_both}
\end{eqnarray} 
We first present the condition for identifying the unobserved $K$. 
\begin{customasp}{FR2} (Full Rank)
    There exists a cutoff of the support so that the distribution of $f^k_{r-1:r-1}(x)$, $k=1,...,K$, where $x \in [\underline x, \xL]$, are linearly independent; and the mixture distribution $\sum_n p_{k,n}  c_{r-1,n}  f^{k}_{1:n-r+1}(y)$, where $y\in [\xL, \bar x]$ and $k=1,...,K$, are linearly independent.
    \label{ass_fullrank2}
\end{customasp}

Following Lemma \ref{id_K}, if Assumption \ref{ass_fullrank2} holds, then the unknown $K$ is identified by the rank of the observed joint distribution of the two consecutive order statistics even if competition is unknown. Therefore, we treat $K$ as known from now on and focus on identifying the state-specific bid/value distributions.

Following Lemmas \ref{id_main}, we can identify state-specific bid functions $f^{k}_{r-1:r-1}(x)$ and $\bar f^k(y) \equiv \sum_n p_{k,n} c_{r-1,n} f^{k}_{1:n-r+1}(y)$ up to different scales and the same ordering, where $x\in l$ and $y\in h$. Consequently, we can identify the state-specific bid distribution $f^{k}(x)$, where $x \le \xL$, but cannot identify it in the ``high" segment because competition is unobserved. To summarize, we identify the state-specific bid distributions up to the same ordering but different scales: 
    \begin{eqnarray*}
    \begin{cases}
   f^{k}(x) & =\eta_l^k \check{f}_l^{k}(x)  \quad  \quad\hbox{if}\hspace{0.2cm} x \leq \xL, \\
   \bar f^k(y) & =\eta_h^k \check{\bar f}_h^{k}(y) \quad   \quad\hbox{if}\hspace{0.2cm} y \geq \xL,
    \end{cases}%
    \label{id_sum2}
    \end{eqnarray*}  
where $\check{f}_l^{k}(\cdot)$ and $ \check{\bar f}_h^{k}(\cdot)$ are the state-specific distributions identified from the above analysis for the ``low" and ``high" segments, respectively. 

To fully identify the state-specific bid  distributions, we need to tackle the following two issues: (1) pin down scales, i.e., $\eta_l^k$ and $\eta_h^k$; (2) identify the state-specific bid distribution for $y \ge \xL$. One possible solution to both issues is to let $\xL$ go to upper bound $\bar x$. However, taking $\xL$ to the limit is at odds with our setting. First, the density of the bottom order statistic $f_{1:n-r+1}^k(y)$, and hence mixture $\bar f^k(y)$, is zero at the limit, which leaves us little data to pin down the scale. Second, our identification strategy relies on the variation of $\bar f^k(y)$ in the $K$ exclusive points/intervals above $\xL$. Specifically, we need sufficient variation in the distribution at those $K$ points/intervals so that matrix $ \boldsymbol{H}$ is invertible and eigendecomposition is feasible. Lastly, the distribution of unobserved factors $p_{k,n}$ is of interest in our setting. Taking $\xL$ to the limit is only useful for identification of the value distributions. 

Instead, we propose addressing the first issue of pinning down the state-specific scales and the probability of the combined unobserved factors, $p_{k,n}$, by plugging the identified marginal densities back into the marginal distributions of the $r^{th}$  order statistic. Specifically, we plug in the identified component distributions with the associated scales, i.e., $\eta_l^k \check{f}_l^{k}(x)$, into the CDF of order statistics $X_{r}$:
\begin{equation} \label{dist2}
F_{r}(x) = \sum_{k,n} p_{k,n} \sum_{i=r}^{n} c_{i,n} [\eta_l^k \check{F}(x)]^{i}[1-\eta_l^k\check{F}(x)]^{n-i}, \forall x \le \xL.
\end{equation}
Intuitively, we can construct polynomial equations with those to-be-identified components as the multivariate unknowns. Altogether, we need to solve for $(K+K\times |n|-1$) unknowns (minus one due to the fact that $\sum_{k,n} p_{k,n}=1$). Note that we can construct a continuum of equations because the bids are continuous. We impose the following assumption to guarantee there exists a unique solution for the scales and weights. 

\begin{customasp}{U}  \label{ass_regu}
The system of equations \eqref{dist2} has a unique solution $\{\eta_l^k, p_{k,n}\}$ in $(0,+\infty)^{K} \times \Gamma$, where $\Gamma = \{(p_{k,n}):p_{k,n} \in (0,1), \sum_{k,n} p_{k,n} =1\}$.\footnote{This is given now that we can use the continuity of the PDF to reduce the number of unknown scales. We can also reconstruct the unknown scales to be the CDF at the cutoff points, which will shrink the interval for scales to be $(0,1)$. }
%{\color{red}In our Online Appendix, we show this condition is trivially satisfied in many cases.}
\end{customasp}

Once the distribution of state and competition $p_{k,n}$ is identified, we can address the second issue of identifying the bid distributions for $y  \ge \xL$ in the following two steps. First, the distribution of the mixture of competition is identified from Equation \eqref{2nd_both} because density $f_{r-1:r-1}(x)$ is identified for $ x \le \xL$, given that scale $\eta_l^k$ is known. Second, we identify the state-specific bid distribution for $y  \ge \xL$ using the fact that the identified state-specific mixture component is a monotonic function of the to-be-identified state-specific bid distribution (see Equation \eqref{eq_mon}).

% Specifically, integrating $\bar f^k(\cdot)$ from $z$ to $\overline x$ leads to the one-to-one mapping:
%\begin{eqnarray}
%\int^{\bar x}_z  \bar f^{k}(z) dz = \sum_n p_{k,n} c_{r,n} \left[ 1-F^k(z) \right]^{n-r+1}, \quad 
%\forall z > \xR, 
%\end{eqnarray}
%where $p_{k,n}$ and $\bar f^{k}(z)$ are identified. The mixture component $\sum_n p_{k,n} c_{r,n} \left[ 1-F^k(z) \right]^{n-r+1}$ is strictly decreasing in  the unknown $F^k(z)$. Therefore, there is a unique solution.
We summarize the identification result in the following theorem.

\begin{theorem}
\label{theorem_kn}
If Assumptions (\ref{ass_fullrank2}), (\ref{ass_distinct}), and (\ref{ass_regu}) are satisfied, then state weight $p_{k,n}$ and the state-specific value distribution $F^{k}(x)$ are identified for ascending IPV auctions with unobservable state and competition using two consecutive order statistics with a binary instrumental variable. 
\end{theorem}

%\paragraph{Remark:} Note that in the situation that the number of bidders is unknown, i.e., $n$ is unknown, getting information of the bottom $r, r-1, r-2$  bids (procurement auction) and the top $r, r-1, r-2$ bids contain different information. Thus, identification differs in these two scenarios. However, the number of bidders enters the joint distribution of the consecutive order statistics in a special way. Specifically, the unknown $n$ only one of the three separate distributions. As a result, we can control the impact of the unknown $n$. As a result, if we know the top $r, r-1, r-2$ bids, we can still identify the underlying type-specific bid distribution. 
\subsection{Discussion about Estimation and Inference}

In this subsection, we briefly discuss a two-step estimation procedure involving estimating $K$ and the state-specific value distributions sequentially.\footnote{Another possibility is to estimate the value distributions jointly with the number of unknown $K$ by MLE that penalizes a larger $K$, which is also out of the scope of this paper.} To estimate $K$, we construct a finite set of discretizations to approximate the continuum counterpart and estate the rank of a general matrix via a sequential test. Once $K$ is estimated, we can estimate the state-specific bid distribution via eigenvalue decomposition or a semiparametric sieve Maximum Likelihood Estimation. We also briefly discuss the consistency of the two-step estimator.

%In this subsection we lay out the main challenges in estimation and inference. Specifically, we discuss briefly how to estimate the cardinality of the unobserved heterogeneity, $K$, and discuss potential directions to leverage sources of over-identification to improve efficiency and discuss how to cope with sequential estimation errors in standard errors. 

\paragraph{Estimation of Cardinality}   
The identification of $K$ suggests an intuitive path to estimate the unknown $K$. We first introduce some notation. Let $D_{\tilde K}$ collect all possible discretizations with the number of intervals being $\tilde K$, let $d$ denote any discretization in the set, and let $\boldsymbol{J}^{d}_{l,h}$ denote the matrix constructed based on discretization $d$. From Lemma \ref{id_K}, we have 
\begin{eqnarray*}
K \ge \tilde K &\text{ if }&  \max_{d \in D_{\tilde K} } \{rank( \boldsymbol{J}^{d}_{l,h})\}=\tilde K, \\
K =\max_{d \in D_{\tilde K} } \{rank( \boldsymbol{J}^{d}_{l,h})\}  & \text{if}& \max_{d \in D_{\tilde K} } \{rank( \boldsymbol{J}^{d}_{l,h})\} <\tilde K. 
\end{eqnarray*}
Therefore, we can start by letting $\tilde K=2$. If $\max_{d \in D_{\tilde K} } \{rank( \boldsymbol{J}^{d}_{l,h})\}=\tilde K$, we should increase $\tilde K$ by 1 and then re-estimate the rank. This whole process continues until the maximum matrix rank stops growing with $\tilde K$. That is, we stop when $\max \{rank( \boldsymbol{J}^{\tilde K}_{l,h})\}=\tilde K-1$ and conclude that $K=\tilde K-1$. 

However, two challenges arise for estimating the unobserved $K$. First, note that the set $D_{\tilde K}$ is continuous, so it is impossible to exhaust the list. Moreover, for identification purposes, we only need to show that there exists one discretization where the full column rank condition holds. However, the identification result is nonconstructive in the sense that it does not provide a straightforward path regarding how to construct such a discretization. One can only try as many discretizations as possible for each $\tilde K$ and hope that we can estimate the $\max_{d \in D_{\tilde K} } \{rank( \boldsymbol{J}^{d}_{l,h})\}$ consistently. The second challenge lies in the estimation of the rank of a general matrix given any discretization. In what follows, we deal with both challenges one by one. 

In practice, we propose approximating the continuous set $D_{\tilde K}$ for each $\tilde K$ using a finite but large set involving the choices of the middle cutoff and grid points in both the $l$ and $h$ segments. We first determine the set of all possible values that the cutoff point $\xL$ can take on. Specifically, suppose we want to try $R_1$ different values of the cutoff $\xL$. Naturally, we should spread out those values in the full support to try to capture its variation, which increases the chance of satisfying the full rank condition. Therefore, we propose to determine these values using the quantiles of the bids. Specifically, we construct the set of cutoff points as $\Omega_{\xL}(R_1) = \{\tau(1/(R_1+1)),....\tau(R_1/(R_1+1))\}$, where $\tau(\cdot)$ indicates the quantile function of the observed bids. Given any cutoff from $\Omega_{\xL}$, we now determine the set of all possible discretizations in each segment by first determining the possible choices of the $\tilde K$ intervals, which is equivalent to choosing $\tilde K-1$ grid points. We denote the set including all possible values of the grid points as $\Omega_{l}$ and $\Omega_{h}$. Let us use segment $l$ as an illustration. Suppose we decide there are $R_l$ possible values that the $\tilde K-1$ grid points can choose from, where $R_l \ge \tilde K-1$. Once again, intuitively, it is  informative to use quantiles to determine these grid points. That is, $\Omega_{l}\equiv \{\tau_l(1/(R_{l}+1)),....,\tau_l(R_{l}/(R_{l}+1))\}$, where $\tau_l(\cdot)$ is the quantile function adjusted based on bids in segment $l$, which generates $C^{R_{l}}_{\tilde K-1}$ possible combinations of the discretization in segment $l$. The same procedure can be conducted for segment $h$. Therefore, we have created $R \equiv R_1\times C^{R_{l}}_{\tilde K-1} \times C^{R_{h}}_{\tilde K-1}$ possible discretizations and denote the approximated set as $\tilde D_{\tilde K} \equiv \{d_r, j=1,...,R\}$, where $d$ denotes any discretization in the set.

\noindent
\textbf{Example: } We now provide a simple example of the construction of $\tilde D_{\tilde K}$. Let $\tilde K=2$. Suppose $R_1=3$, indicating that we would like to try three different cutoff points: quantiles 0.25, 0.5, 0.75, respectively (i.e., $\Omega_{\xL}(R_1=3) =\{\tau(0.25), \tau(0.5), \tau(0.75)\}$). We use the median cutoff as an example for discretization of the segment $l$. Since $\tilde K=2$, we only need to determine one grid point, which will produce two exclusive intervals. Specifically, suppose we allow three different values of the grid points, $R_l=R_h=3$, so $\Omega_l=\{\tau_l(0.25), \tau_l(0.5), \tau_l(0.75)\}=\{\tau(0.125), \tau(0.25), \tau(0.375)\}$, and $\Omega_h=\{\tau_h(0.25), \tau_h(0.5), \tau_h(0.75)\}=\{\tau(0.625), \tau(0.75), \tau(0.875)\}$. Therefore, we have constructed the approximation of the original continuum set as $\tilde D_{\tilde K}=\{(i,j,k)\}_{i,j,k}$, where $i$ refers to the middle cutoff, and $j,k$ the grid point chosen for segment $l$ and $h$, respectively. The cardinality of the 
constructed set can be computed as $R=3*3*3=27$, because  each $i,j,k$ can be chosen from three possible values.

%First of all, we propose to pick a maximum numbers of discretizations $J_{\tilde K}$ for each $\tilde K$, which set to be reasonable large and can depend on $\tilde K$. Secondly, given  $J_{\tilde K}$,  we propose in the follows how to decide specific discretizations in the set, which involves deciding the cutoff point $\xL$ and the grid points in each segment.  Specifically, To pick the cutoff point $\xL$, it might be useful to first divide the support by the median of the bids so that the number of observations in both segments are balancing, which is important for estimating the joint probability matrix with enough accuracy. That is, one can first divide the support as $l \equiv [\underline x, \xL=median(x)]$ and $h \equiv [\xL=median(x), \bar x]$. Given the cutoff and the chosen number of intervals $\tilde K$, similarly we might divide two segments based on the quantiles of the bids in each segment so that each bins has a reasonable sample size. 

We now discuss how to estimate the rank of matrix $ \boldsymbol{J}^{d}_{l,h}$ for a given discretization $d$. Specifically, we first estimate every element in the joint probability matrix with $\tilde K$ intervals in each segment using a simple frequency estimator. That is,
\begin{eqnarray}  \notag
   \boldsymbol{ \hat  J}^{d}_{l,h}= \left(\frac{1}{M} \sum^M_{m=1} I(X_{r-1:n} \in l_i, X_{r:n} \in h_j)\right)_{i,j}.
\end{eqnarray}

Given the probability matrix, we estimate the rank of matrix $ \boldsymbol{ J}^{d}_{l,h}$ using a sequence of tests, following \cite{robin2000tests}. Specifically, we construct the hypotheses as: $H^r_0: rank(\boldsymbol{J}^{d}_{l,h})=r$  against the alternatives $H^r_1: rank(\boldsymbol{J}^{d}_{l,h}) > r$ with $r=1,2,...,\tilde  K$. The sequence of tests proceeds as follows. First, we start with the null hypothesis that the rank of matrix $\boldsymbol{J}^{d}_{l,h}$ is 1. If such a null hypothesis is rejected,  we augment $r$ by one and repeat the test.  When we fail to reject the null of $rank(\boldsymbol{J}^{d}_{l,h})=r$ for the first time,  the rank of $\boldsymbol{J}^{d}_{l,h}$ is estimated as $r$. 

%We can summarize the estimation of the rank for $\tilde K$ intervals as
%\begin{eqnarray}
%\hat r_{\tilde K}=max_{d_j \in \tilde D_{\tilde K}}\{rank(   \boldsymbol{  J}^{d_j}_{l,h})\}.
%\end{eqnarray}

%If we fail to reject that matrix $\boldsymbol{J}^{\tilde K_j}_{l,h}$ is full rank, i.e., $rank (\boldsymbol{J}^{\tilde K_j}_{l,h})= \tilde K$, it suggests that at least $\tilde K$ underlying distribution is linearly independent. So for the first time of the experiment with $\tilde K$ that matrix $\boldsymbol{J}^{\tilde K_j}_{l,h}$ is full rank, one can stop experimenting for such a $\tilde K$, and should increase $\tilde K$ to be $\tilde K+1$. However, if the rank of $\boldsymbol{J}^{\tilde K_j}_{l,h}$ is estimated to be smaller than $\tilde K$, we know from the identification that the unknown $K$ is at least $\boldsymbol{J}^{\tilde K_j}_{l,h}$, but it does not guarantee that the full rank condition is satisfied for such a discretization. Therefore, one should experiment new discretizations and repeat the above procedure until either the maximum number of experiments $J$ is reached or we fail to reject that matrix $\boldsymbol{J}^{\tilde K_j}_{l,h}$ is full rank, so we need to increase $\tilde K$ to be $\tilde K+1$. 
Therefore, the unknown $K$ can be estimated as
\begin{eqnarray}
\hat K=\max_{\tilde K}\{ \max_{d_j \in \tilde D_{\tilde K}}\{ \widehat {rank(\boldsymbol{  J}^{d_j}_{l,h}) } \} \}.
\end{eqnarray}

Note that such an estimator does not require an optimization algorithm, so it is easy to compute. Such an estimator is unconventional, and the consistency of such an estimator is not trivial. Two issues need to be resolved for consistency. The first issue is determining how to guarantee the rank estimator is consistent. This problem is well studied in the existing literature (see \cite{robin2000tests}). The consistency of such an estimator relies on the selection of the significance levels for the sequential tests. Specifically, as sample size $M$ increases, significance level $\alpha_{M}$ should go to infinity but at a lower rate. The second issue is ascertaining how well our proposed simplified set $\tilde D_{\tilde K}$ approximates the set of all possible discretization $D_{\tilde K}$. The answer to this question depends on the specification of the discretization. Intuitively, the more experiments we try, i.e., the larger of $R_1$, $R_{l}$, and $R_{h}$, the better the approximation is. Note that given one discretization, we can estimate the rank fairly quickly, since no optimization algorithm is needed. Thus, in practice, one could try very large $R_1$, $R_{l}$, and $R_{h}$. Intuitively, if $R_1$, $R_{l}$, and $R_{h}$ go to infinite as the sample size goes to infinity, $\tilde D_{\tilde K}$ should approximate the true $D_{\tilde K}$ very well. However, establishing such a theoretical result is outside the scope of this paper, so we leave it for future research. Note that the estimation of $K$ serves as a model selection procedure. Therefore, ``if the selection is consistent, i.e., $prob(\hat K_n=K) \rightarrow 1$ as $M \rightarrow \infty$, the asymptotic properties of any statistic based on the true model and the selected model are identical and hence asymptotic inference is unaffected,"(Lemma 1, \cite{potscher1991effects}).

%The sconwe adjust the asymptotic size $\alpha_r$ of the CR test at each stage $r$ to depend on the sample size $M$  with a certain rate. The revised critical region at stage $r$ is given by $\{\mathcal {CRT}_r \ge c^r_{1-\alpha_{r M}}\}$ with the critical value $c^r_{1-\alpha_{r M}}$ along with an asymptotic size $\alpha_{rM}$ under the null $H^r_0: Rank(J_{l,m_1,h})=r, r=1,2,...,K$. The estimator for the cardinality of the support of the UH $K$ then is defined as 
%\begin{eqnarray*}
%\hat r \equiv min_{r \in \{1,...,K\}} \{r: \mathcal {CRT}_r \ge c^i_{1-\alpha_{iM}}, i=1,...,r-1, \mathcal {CRT}_r < c^r_{1-\alpha_{rM}}\}.
%\end{eqnarray*}

% We summarize the condition for consistency of the rank estimator $\hat K$ as follows:
%
%\begin{lemma} The estimator of  $rank(\boldsymbol{ \hat  J}^{\tilde K}_{l,h})$, i.e., $r$, is weakely consistent, that is, $r \rightarrow_p rank(\boldsymbol{ \hat  J}^{\tilde K}_{l,h})$, if the asymptotic size for the sequential tests satisfies the following conditions. (i), $\alpha_{M} \rightarrow 0$; (ii), $lim_{M\rightarrow\infty}  \frac{ ln \alpha_{M}}{M}=0$.
%\end{lemma}
%\begin{proof}
%See \cite{robin2000tests}.
%\end{proof}

\paragraph{Estimation of the State-specific Value Distribution} Once $K$ is estimated, we can estimate the state-specific bid/value distribution, given the discretization,  by following the identification steps one by one since it is constructive.  Suppose there is a subset of $R$ experiments where the rank of the joint probability matrix is the same as $\hat K$. This suggests that the model is over-identified. We can estimate the state-specific bid/value distribution by every discretization with the joint probability matrix being full rank. In this case, we can leverage such over-identification power and improve estimation efficiency by averaging over all those estimators. 

However, estimation following the identification strategy involves multiple steps, with the eigendecomposition being the first step. Once the decomposition is achieved, we need to estimate the density of the order statistics,  then invoke the one-to-one mapping to estimate the parent distribution from the density of its order statistic, and then pin down the scales. Such a sequential estimation procedure is usually inefficient. Therefore, we propose to directly estimate the state-specific bid/value distribution  using a semi-parametric sieve estimator. Specifically, we can approximate each state-specific value distribution using a sieve base. Therefore, we just need to estimate $\hat K \times L$ sieve coefficients, where $L$ is the number of base functions chosen. Specifically, 
we first introduce the Bernstein series for semiparametrically estimating the underlying state-specific bid distribution $f^k(x)$. We use the following specification  to approximate the unknown bid distribution, which is the value distribution in the scenario of ascending auctions. That is,
\begin{eqnarray} \notag
\check f^k(x;b_k) \equiv \sum^{L_k}_{l=1}b_{lk}  beta(x;l,L_k-l+1) \simeq f^k(x),
\end{eqnarray}
where $L_k$ is a smoothing parameter that increases with sample size, and $b_k \equiv \{b_{1k},...,b_{L_kk}\}$ is the vector of sieve parameters for state $k$ to be estimated. As $ f^k(x;b_k)$ is a density function, it has to be non-negative, and its integration over the domain $[0,1]$ is 1. That is, $b_{1k} \ge 0$ and $\sum_l b_{lk}=1$. The CDF of the state-specific bid distribution  can then be approximated as:
\[\check F^k(x;b_k) \equiv \int^x_{0} \check f^k(x;b_k)d x \simeq \int^x_{0} f^k(x)d x =F^k(x).\]

Let $\theta_{sieve}$ collect all sieve parameters, i.e., $\theta_{sieve} \equiv \{b_{1k},...,b_{L_kk}, p_k, \Pr(W|k)\}_{k}$, where $k=1,...,K$. The semi-parametric sieve estimator $\hat \theta_{sieve}$ is the one that maximizes the log likelihood of the joint distribution of the two observed order statistics and the instrumental variable. That is, 
\begin{eqnarray} \notag
\widehat \theta_{sieve} &=& \arg \max_{\theta} \sum_m \log \big[ \sum_k p_k \check f^k_{r-1,r} (x^m_{r-1},x^m_r,W;b_k)\big]. 
%&=& argmax_{\theta} \sum_m \sum_k \lambda^{t}_{mk} \Big[ (r-3)ln\check F^k(x^m_{r-2}:\theta) + ln \check f^k(x^m_{r-2}:\theta) +   ln \check f^k(x^m_{r-1}:\theta) \notag \\ 
%&& \quad \quad \quad \quad \quad \quad \quad  \quad \quad  + (n-r) ln(1-\check F^k(x^m_r:\theta)+ ln \check f^k(x^m_r:\theta). \Big]
\end{eqnarray}
\cite{ghosal2001convergence} proves that the convergence for semi-parametric sieve MLE with Bernstein polynomial base functions is at ``nearly parametric rate" $\sqrt {\frac{\log M}{M}}$ for Hellinger distance, when the true density is indeed a Bernstein density. When the true density is not of the Bernstein type, the sieve estimator converges at rate $(\frac{\log M}{M})^{1/3}$ if the true density is twice differentiable and bounded away from 0. 

%Another possible direction is to estimate the state-specific bid/value distribution jointly with the number of unknown $K$ by maximum likelihood estimation but penalize a larger $K$, which is also out of the scope of this paper.

    \section{Conclusion}
    
Auction data often fail to record all bids or all relevant auction-specific characteristics that shift bidder values. Instead, they may contain only a subset or order statistics of the bids and suffer from unobserved heterogeneity (UH). In this paper, we present a set of new identification results for auction models with discrete UH using consecutive order statistics. In particular, we show that despite correlation between order statistics, employing the same number of measurements is sufficient for achieving identification. 

Mixture models for UH usually rely on independence assumptions. Our results show that exploring the statistical/model structure is a fruitful approach to restore identification when independence fails. While this paper focuses on IPV auction models and finite UH, natural extensions include affiliated private value and continuous UH, as seen in \citet{li2002structural}, \citet{LV1998}, and \citet{K2011}. Moreover, UH and data truncation arise in other settings, such as beauty contests and war of attrition models, where many players compete for multiple prizes. We leave these for future research.
    
    %Note that consecutiveness plays a key role in the identification of nonseparable finite unobserved heterogeneity without support variations. A possible extension to the identification results is to explore the markov property of order statistics and investigate the possible identification results using any four order statistics. The identification strategy is similar to some recent results on identification of dynamic models with unobserved state variables (\cite{hu2012nonparametric} and \cite{Luoxiao2018}). Specifically, first, the joint distribution of four order statistics can be represented as a multiplicatively separable mixture structure by the Markov property, with which an eigenvalue decomposition argument identifies a key matrix that governs the finite mixture structure. Second, we apply the matrix identified in the first step to identify the component distributions in the lower portion of the support and then in the upper portion using joint distribution of only three order statistics. Moreover, the same identification argument applies to the scenario of nonseparable continuous unobserved heterogeneity.
    
    %See also \citet{M2017} 
    
    \newpage 

%\hspace{6cm} \textbf{{\LARGE {Appendix}}} 

\appendix

%\numberwithin{equation}{section} \renewcommand\theequation{\Alph{section}.%
%\arabic{equation}}
\setcounter{equation}{0}
\numberwithin{equation}{section} \renewcommand\thesection{A}
\renewcommand\thesection{A}    

\section*{Appendix}

\subsection{Proof of Lemma \ref{cOS}}

Note that $f_{1:n}(x) = n[1-F(x)]^{n-1}f(x)$ and $f_{n:n}(x) = n[F(x)]^{n-1}f(x)$. We have  
\begin{align*}
f_{r-1,r:n} (x,y)  
& = \frac{n!}{(r-2)!(n-r)!} [F(x)]^{r-2} f(x) [1-F(y)]^{n-r} f(y) \\
& =  \frac{n!}{(r-2)!(n-r)!} \frac{f_{r-1:r-1}(x)}{r-1} \frac{f_{1:n-r+1}(y)}{n-r+1} \\
& =  \frac{n!}{(r-1)!(n-r+1)!} f_{r-1:r-1}(x) f_{1:n-r+1}(y) .
\end{align*}

\subsection{Proof of Lemma \ref{id_K}: Identification of the Unknown K} \label{idK} Note that the joint probability matrix of the two order statistics with any discretization $\tilde{K}$ can be represented as 
    \begin{eqnarray}
    \boldsymbol{J}^{\tilde K}_{l,h}= \boldsymbol{L}^{\tilde K} \boldsymbol{D}_p {\boldsymbol{H}^{\tilde K}}^T,
    \end{eqnarray}
where the observed matrix $\boldsymbol{J}^{\tilde K}_{l,h}$ has dimensions $\tilde K \times \tilde K$, the matrices $ \boldsymbol{L}^{\tilde K} $ and $\boldsymbol{H}^{\tilde K}$ have dimensions $\tilde K \times K$, and $\boldsymbol{D}_p$ are diagonal matrices with dimensions of $ K \times  K$.

%\begin{lemma}\label{id_K} The rank of the joint probability matrix provides the following information about the unknown $K$.\\
%1). With any $\tilde K$-interval discretization, $Rank(\boldsymbol{J}^{\tilde K}_{l,h}) \le K$.\\
%2). If  Assumption \ref{ass_fullrank1} holds, there exists a $\tilde K$-interval discretization, where $\tilde K \ge K$, such that $Rank(\boldsymbol{J}^{\tilde K}_{l,h}) = K$. 
%\end{lemma}
We first introduce the following two properties of the rank inequality. First, 
\begin{eqnarray}
Rank(\boldsymbol{J}^{\tilde K}_{l,h} )&=& Rank(\boldsymbol{L}^{\tilde K}  \boldsymbol{D}_p {\boldsymbol{H}^{\tilde K}}^T) \le min \{Rank(\boldsymbol{L}^{\tilde K} ), Rank(\boldsymbol{H}^{\tilde K})\},
\label{id_rank1}
\end{eqnarray}
which trivially holds for any number of intervals and imposes no restrictions on the model primitives. Second, by Sylvester's rank inequality, we have
\begin{eqnarray}
Rank(\boldsymbol{J}^{\tilde K}_{l,h} )&=& Rank(\boldsymbol{L}^{\tilde K}  \boldsymbol{D}_p {\boldsymbol{H}^{\tilde K}}^T) \ge Rank(\boldsymbol{L}^{\tilde K} \boldsymbol{D}_p) + Rank({\boldsymbol{H}^{\tilde K}}) -min\{K,\tilde K\}.\notag \\
\label{id_rank2}
\end{eqnarray} 
If Assumption (\ref{ass_fullrank1}) holds, for any $\tilde K \le K$, there exists at least one discretization of $\tilde K$ intervals so that the unobserved matrix is full row rank, i.e., $Rank(\boldsymbol{L}^{\tilde K} )=Rank(\boldsymbol{H}^{\tilde K})=\tilde K$. Therefore, we have 
\begin{eqnarray}
\tilde K=Rank(\boldsymbol{J}^{\tilde K}_{l,h}), 
\end{eqnarray}
because $Rank(\boldsymbol{J}^{\tilde K}_{l,h} )\le \tilde K$ by Equation \eqref{id_rank1}, and $Rank(\boldsymbol{J}^{\tilde K}_{l,h} )\ge \tilde K$ by Equation \eqref{id_rank2}.

Moreover, if $\tilde K \ge  K$ and Assumption (\ref{ass_fullrank1}) holds, there exists a discretization of $\tilde K$ intervals where both matrices $\boldsymbol{L}^{{\tilde K}}$ and and $\boldsymbol{H}^{{\tilde K}}$ are full column rank, i.e., $Rank(\boldsymbol{L}^{\tilde K})=Rank(\boldsymbol{H}^{\tilde K})=K$.  Therefore, we can show that $K$ can be identified as the rank of the probability matrix, i.e., 
\begin{eqnarray}
K=Rank(\boldsymbol{J}^{\tilde K}_{l,h} ),
\end{eqnarray}
because $Rank(\boldsymbol{J}^{\tilde K}_{l,h} )\le K$ by Equation \eqref{id_rank1}, and $Rank(\boldsymbol{J}^{\tilde K}_{l,h} )\ge K$ by Equation \eqref{id_rank2}.\footnote{A similar identification approach is exploited in  \citet{an2017identification} and \cite{xiao2014identification}.} 

\subsection{Proof of Theorem 1}
 
 We provide proofs for Lemmas \ref{id_main} and \ref{lemma_scales}, which constitute the proof for Theorem 1. 
    \paragraph{Proof of Lemma \ref{id_main}\quad Step 1: Identification in the ``low" segment}

The relevance assumption of the instrumental variable leads to the following main equation for identification:
    \begin{eqnarray*}
    \boldsymbol{J}_{l,h,0} \boldsymbol{J}^{-1}_{l,h}= \boldsymbol{L}   \boldsymbol{D}_{0} \boldsymbol{L}^{-1} ,
    \end{eqnarray*}
    where $\boldsymbol{D}_{0}$ is a diagonal matrix with the $k^{th}$ diagonal element as $Pr(W=0|k)$. With the relevant assumption of the instrument variable, the eigendecomposition is unique so that we can uniquely identify the probability matrix $\boldsymbol{L}$ as the eigenvector matrix through an eigenvalue and eigenvector decomposition of the observed matrix $ \boldsymbol{J}_{l,h,0} \boldsymbol{J}^{-1}_{l,h}$. Note that such identification is up to permutations and scales.
 
     After identifying probability matrix $\boldsymbol{L}$, we further identify the density of order statistic $X^k_{r-1:r-1}$ for any value in this segment. We then use the one-to-one mapping between an order statistic's density and its parent distribution to recover the state-specific density distribution in the ``low'' segment. In particular, to identify the density for order statistic $X^k_{r-1:r-1}$, we again use the joint distribution. Note that the matrix representation in Equation (\ref{eq_joint_M1}) also holds for a particular value of $x \in l$:
    \begin{eqnarray}
    \label{eq_joint2}
    \boldsymbol{J}_{x,h}= \boldsymbol{L}_x  \boldsymbol{D_p} \boldsymbol{H}^T,
    \end{eqnarray}
    where $\boldsymbol{J}_{x,h}  \equiv \{ \int_{y \in h_j} f_{r-1,r:n}( x,y) d y \}_j $ and $\boldsymbol{L}_x = \{f^k_{r-1:r-1}(x)\}_k$ are $1\times K$ row vectors representing the counterparts of matrices $\boldsymbol{J}_{l,h}$ and $\boldsymbol{L}$, respectively, by replacing the $K$ intervals with a particular value of $x \in l$. Therefore, $\boldsymbol{J}_{x,h}$ and $\boldsymbol{L}_x$ represent PDFs instead of probabilities. Note that both Equations (\ref{eq_joint_M1}) and (\ref{eq_joint2}) have the common component $\boldsymbol{D_p} \boldsymbol{H}^T$, which is invertible. As a result, we can identify the state-specific density for order statistic $X^k_{r-1:r-1}$ in the ``low" segment, i.e., $f^k_{r-1:r-1}(x),\forall x\in l$, as in the following closed-form expression:
    \begin{eqnarray*}
    \label{id_l1}
    \boldsymbol{L}_x=\boldsymbol{J}_{x,h}\boldsymbol{J}^{-1}_{l,h} \boldsymbol{L}.
    \end{eqnarray*}
Density vector $\boldsymbol{L}_x$ is identified up to the same scales and ordering as that of probability matrix $\boldsymbol{L}$. Note that the components identified from the eigenvalue-eigenvector decomposition are not the probability matrix $\boldsymbol{L} $ itself, but a scalar multiplication of each of its columns with an unknown constant. That is, denote the eigenvector matrix obtained from the decomposition $ \boldsymbol{\check L} $. We have $ \boldsymbol{L}=  \boldsymbol{\check L}   \boldsymbol{\lambda}_l$, where $ \boldsymbol{\lambda }_l\equiv diag[\lambda^1_l,...,\lambda^K_l]$ is the scale matrix. 
    Based on  matrix $\boldsymbol{\check L}$ from the decomposition, we can compute the corresponding density vector as follows:
    \begin{eqnarray}
    \boldsymbol{\check L}_x &\equiv & J_{x,h} J_{l,h}^{-1} \boldsymbol{\check L}\notag \\
     &=& J_{x, h} J_{l,h}^{-1} \boldsymbol{\check L}  \boldsymbol{\lambda}_l  \boldsymbol{ \lambda}_l^{-1}\notag \\
      &=& J_{x, h} J_{l,h}^{-1} \boldsymbol{L} \boldsymbol{\lambda}_l^{-1} \notag \\
        &=& \boldsymbol{L}_x \boldsymbol{\lambda}_l^{-1}. \notag \\
    \text{Thus,} \quad \quad \quad \boldsymbol{L}_x&=&    \boldsymbol{\check L}_x \boldsymbol{\lambda}_l.
    \end{eqnarray}
    As a result, the type-specific density for order statistics $X^k_{r-1:r-1}$ in the ``low" portion is identified up to the same scale as the probability matrix  $ \boldsymbol{L}$, i.e., $f^k_{r-1:r-1}(x)=\lambda^k_l \check f^k_{r-1:r-1}(x)$, where  $\check f^k_{r-1:r-1}(x)$ is the $k^{th}$ element in vector $\boldsymbol{\check L}_x$.
    
    Next, we show that state-specific density $f^k(x)$ is also identified up to scales as follows:  
    \begin{eqnarray}
    f^k_{r-1:r-1}(x) &=& (r-1)  [F^k(x)]^{r-2} f^k(x) \notag \\
    \leftrightarrow \quad  \int^{x}_{\underline x}  f^k_{r-1:r-1}(v) d v &=&  (r-1) \int^{x}_{\underline x} [F^k(v)]^{r-2} f^k(v) d v \notag\\
 \leftrightarrow \quad  \int^{x}_{\underline x}  f^k_{r-1:r-1}(v)d v &=& [F^k(x)]^{r-1} \notag\\
    \leftrightarrow F^k(x) &=&   \Big[\int_{\underline x}^{y}  f^k_{r-1:r-1}(v)dv \Big]^{\frac{1}{r-1}}, \notag 
    \end{eqnarray}
    where the first equality holds due to the definition of order statistics, and the second equality holds by taking the integration on the right-hand size and $F^k(\underline x)=0$.
Therefore, we can derive the state-specific density for $x\in l$ as follows:
    \begin{eqnarray}
    f^k(x) &=&  \frac{1}{r-1} \Big[\int_{\underline x}^{x}  f^k_{r-1:r-1}(v)d v \Big]^{\frac{1}{r-1}-1} f^k_{r-1:r-1}(x) \notag \\
    &=& \frac{1}{r-1} \Big[\int_{\underline x}^{x} \lambda^k_l \check f^k_{r-1:r-1}(v) d v \Big]^{\frac{1}{r-1}-1} \lambda^k_l \check f^k_{r-1:r-1}(y) \notag \\
    &=&\left(\lambda^k_l\right)^{\frac{1}{r-1}}  \frac{1}{r-1} \Big[\int_{\underline x}^{x} \check f^k_{r-1:r-1}(v) d v \Big]^{\frac{1}{r-1}-1}\check f^k_{r-1:r-1}(x) \notag\\
%    &\equiv &\left(\lambda^k_l\right)^{\frac{1}{r-2}}  \check f_l^k(x),\notag \\
    &\equiv & \eta^k_l  \check f_l^k(x),
    \end{eqnarray}
    where  $\check f_l^k(x) \equiv \frac{1}{r-1} \Big[\int_{\underline x}^{x} \check f^k_{r-1:r-1}(v) d v \Big]^{\frac{1}{r-1}-1}\check f^k_{r-1:r-1}(x)$ represents the state-specific density computed using the eigenvector matrix directly from the decomposition, which is known, and $\eta^k_l\equiv \left(\lambda^k_l\right)^{\frac{1}{r-1}}$ is the scale for the component density. The state-specific density in the ``low" portion is identified up to scale of $\left(\lambda^k_l\right)^{\frac{1}{r-1}}$.

%    Now, we proceed to identify the state-specific parent density function using one-to-one mapping between this density and the identified distribution of its order statistic in the ``low" segment. In particular, the parent density function can be represented as a closed-form function of the identified distribution of its order statistics:
%    \begin{eqnarray}
%    \label{id_F1}
%    f^k(x) &=&  \frac{1}{r-2} \Big[\int_{\underline x}^{x}  f^k_{r-2:r-2}(v)d v \Big]^{\frac{1}{r-2}-1} f^k_{r-2:r-2}(x), \quad \forall x\in l.
%    \end{eqnarray}
%    Again, the state-specific density is identified up to scales since the state-specific density of order statistic $X^k_{r-2:r-2}$ is identified up to scales. Note that the two scales may differ because $   f^k(x)$ is not linear in $f^k_{r-2:r-2}(x)$. Moreover, the state-specific density is identified up to the same ordering of probability matrix $\boldsymbol{L}$. 

%Similarly, in segment ``high", $f^k(z)$ is identified up to the same permutation but different scales. We summarize the above results in the following lemma and leave the details of the proof to the Appendix. 

\noindent
\textbf{Step 2: Identification in the ``high" segment}

In what follows, we identify the state-specific density function $f^k(y)$ in the ``high" segment of the support. We do this by replacing the intervals in $h$ with a particular value $y \in h$ and use matrix invertibility to cancel out common components. Then, we recover the state-specific density function using its relationship with the state-specific density of order statistics. Note that $f^k_{1:n-r+1}(y) $ is identified up to scales, so the state-specific density $f^k(y), \forall y\in h$, is also identified up to scales. 

%Again, the scales may differ. 

%In particular,     \begin{eqnarray}    f^k(z) &=& \frac{1}{n-r+1} \Big[\int_{z}^{\bar x}  f^k_{1:n-r+1}(v) d v \Big]^{\frac{1}{n-r+1}-1}f^k_{1:n-r+1}(z) , \quad \forall z\in h.    \label{eq_h}  \end{eqnarray}    

    First, we can identify probability matrix $\boldsymbol{H}$ up to scales from Equation (\ref{eq_joint_M1}) as the following closed-form expression:
    \begin{eqnarray}
    \boldsymbol{H}^T &=& \left[ \boldsymbol{L}  \boldsymbol{D}_{p} \right]^{-1} \boldsymbol{J}_{l,h} \notag \\
    &=& \boldsymbol{D}^{-1}_p \left[ \boldsymbol{\check L}   \boldsymbol{\lambda}_l  \right]^{-1} \boldsymbol{J}_{l,h} \notag\\
    &=& \boldsymbol{\lambda}_l^{-1} \boldsymbol{D}^{-1}_p\left[ \boldsymbol{\check L}   \right]^{-1} \boldsymbol{J}_{l,h} \notag\\
    &\equiv& \boldsymbol{\lambda}_h \boldsymbol{\check H}^T,
    \end{eqnarray}
    where $\boldsymbol{\lambda}_h \equiv \boldsymbol{\lambda}_l^{-1} \boldsymbol{D}^{-1}_p  \equiv diag \{\lambda^1_h,...,\lambda^K_h\}$ is a diagonal matrix that captures the unknown scales, and $\boldsymbol{\check H}^T$ represents the component that can be computed using results from the decomposition. Since $ \boldsymbol{L}$ is identified up to scales due to the decomposition, and  $\boldsymbol{D}_p$ are diagonal matrices, $\boldsymbol{H}$ can be identified up to scales, but the scales are different from the scales in $ \boldsymbol{L}$.
    
    Using the same logic in the case of the ``low" segment, we can identify the state-specific density for the ``high" segment as follows. Specifically, we obtain: 
    \begin{eqnarray}
    \boldsymbol{H}_y^T 
%    &=& \boldsymbol{\lambda}^{-1}_h \boldsymbol{D}^{-1}_p \boldsymbol{D}^{-1}_{m_{1}} \left[c_{r,n} \boldsymbol{\check L}_x   \right]^{-1} \boldsymbol{J}_{l,m_{1},h} \notag\\
    &= & \boldsymbol{\lambda}_h \boldsymbol{\check H}_y^T,
    \end{eqnarray}
    where $\boldsymbol{H}_y^T$ and $\boldsymbol{\check H}_y^T$ are defined similarly to $\boldsymbol{H}^T$ and $\boldsymbol{\check H}^T$, respectively. This indicates that the state-specific density in the ``high" portion can be identified up to scales; i.e., $f^k_{1:n-r+1}(y) = \lambda^k_h  \check f^k_{1:n-r+1}(y)$, where $\check f^k_{1:n-r+1}(y)$ is the $k^{th}$ component in  vector $\boldsymbol{\check H}_y^T$, which can be computed using results from the decomposition. 
    
    Next, we recover state-specific density using its relationship with the state-specific density of its order statistics. In particular,  by definition of the order statistics, for any value in the ``high" portion, i.e., $y \in h$, we have:
    \begin{eqnarray}
    f^k_{1:n-r+1}(y) &=& (n-r+1)  [1-F^k(y)]^{n-r} f^k(y) \notag \\
    \leftrightarrow \quad  \int^{\bar x}_{y}  f^k_{1:n-r+1}(x) d x &=& (n-r+1) \int^{\bar x}_{y}  [1-F^k(x)]^{n-r} f^k(x)d x \notag\\
    %&=&- \frac{1}{n-r+1}  [1-F^k(x)]^{n-r+1} \mid^{x}_{\underline x}  \\
    \leftrightarrow \quad  \int^{\bar x}_{y}  f^k_{1:n-r+1}(x) d x &=&[1-F^k(y)]^{n-r+1} \notag\\
    %1-[1-F^k(x)]^{n-r+1} &=& \frac{\int^{x}_{\underline x} g^k(x_r) d x_r}{\int^{\bar x}_{\underline x} g^k(x_r) d x_r} \notag\\
    %\leftrightarrow [1-F^k(x)]^{n-r+1} &=& \frac{\int^{\bar x}_{x} g^k(x_r) d x_r}{\int^{\bar x}_{\underline x} g^k(x_r) d x_r} \notag\\
    \leftrightarrow F^k(y) &=& 1-  \Big[\int_{y}^{\bar x}  f^k_{1:n-r+1}(x) d x \Big]^{\frac{1}{n-r+1}}, \notag
    \end{eqnarray}
    where the first equality holds by definition, the second one holds by only taking the integral from $y \ge \xL$ up to the upper bound because we can only identify the density in the high segment. As a result, we can link the state-specific density to the state-specific density of the order statistics as follows, $\forall y\in h$:
    \begin{eqnarray}
    f^k(y) &=& \frac{1}{n-r+1} \Big[\int_{y}^{\bar x}  f^k_{1:n-r+1}(v) d v \Big]^{\frac{1}{n-r+1}-1}f^k_{1:n-r+1}(y)  \notag \\
    &=& \frac{1}{n-r+1} \Big[\int_{y}^{\bar x}  \lambda^k_h \check f^k_{1:n-r+1}(v) d v \Big]^{\frac{1}{n-r+1}-1}\lambda^k_h  \check f^k_{1:n-r+1}(y)  \notag \\
    &=&\left(\lambda^k_h\right)^{ \frac{1}{n-r+1}}  \Big[\int_{y}^{\bar x} \check f^k_{1:n-r+1}(v) d v \Big]^{\frac{1}{n-r+1}-1} \check f^k_{1:n-r+1}(y)  \notag \\
%    &\equiv & \left(\lambda^k_h\right)^{ \frac{1}{n-r+1}}  \boldsymbol{\check f}_h^k(x),
    & \equiv &  \eta^k_h  \check  f_h^k(y),
    \end{eqnarray}
    where $\boldsymbol{\check f}_h^k(y) \equiv \Big[\int_{y}^{\bar x} \check f^k_{1:n-r+1}(v) d v \Big]^{\frac{1}{n-r+1}-1} \check f^k_{1:n-r+1}(y)$, which can be computed directly, and $\eta^k_h\equiv \left(\lambda^k_h\right)^{ \frac{1}{n-r+1}}$ is the unknown scale for the component density in the ``high" segment. The state-specific density in the ``high" portion is identified up to scale of $\left(\lambda^k_h\right)^{ \frac{1}{n-r+1}} $.

    \paragraph{Proof of Lemma \ref{lemma_scales}}  To pin down both scale parameters for each state $k$, we first summarize the identified state-specific density from both segments as follows:
    \begin{eqnarray*}
    f^k(x) =
    \begin{cases}
 \eta^k_l  \check f_l^k(x), \quad  \quad\hbox{if}\hspace{0.2cm} x \in l=[\underline x, \xL], \\
%    \eta^k_m \check  f^k_m(x), \quad  \quad \hbox{if}\hspace{0.2cm} x\in m=[\xL, \xR], \\
 \eta^k_h  \check  f_h^k(x),\quad \quad  \hbox{if}\hspace{0.2cm} x\in h=[\xL,\bar x].
    \end{cases}%
    \end{eqnarray*}
    
    Note that we identify the state-specific density $ f^k(x)$ for the ``low" and ``high" segments up to different scales. Two conditions are required to precisely pin down both scales, i.e., $\eta^k_l$ and $\eta^k_h$. First, the two separately identified density functions should conjoin such that, at the cutoff point $\xL$ , the densities on either side meet at the same value:
    \begin{eqnarray*}
     \eta^k_l  \check f_l^k(\xL) &=& \eta^k_h \check  f^k_h(\xL).
%    \eta^k_m \check  f^k_m(\xR) &=&  \eta^k_h  \check  f_h^k(\xR).
    \end{eqnarray*}
    Moreover, the density over the full support should integrate to 1, which provides the third restriction on the scales:
    \begin{eqnarray*}
    &&\int_{x\in l}      \eta^k_l  \check f_l^k(x)dx + \int_{x\in h}     \eta^k_l  \check f_h^k(x) dx=1.
%    &&\left(\lambda^k_l\right)^{\frac{1}{r-2}}  \int_{x\in l}  \check f_l^k(x)dx+ \lambda^k_m \int_{x\in m}  \check f^k_m(x)dx+ \left(\lambda^k_h\right)^{ \frac{1}{n-r+1}}  \int_{x\in h}  \boldsymbol{\check f}_h^k(x) dx=1.
    \end{eqnarray*}
    The above two conditions lead to the following linear system of equations, in which only the scales are unknowns:
    \[
    \begin{bmatrix}
 \check f_l^k(\xL)       &  -\check f_h^k(\xL)    \\

     \int_{x\in l}  \check f_l^k(x)dx &  \int_{x\in h}  \check f_h^k(x)dx
    \end{bmatrix}
    \begin{bmatrix}
    \eta^k_l\\
    \eta^k_h 
    \end{bmatrix} =  \begin{bmatrix}
    0\\
    1
    \end{bmatrix}.
    \]
    Pinning down the scales requires the 
    matrix $$M\equiv \begin{bmatrix}
    \check f_l^k(\xL)        & - \check f^k_h(\xL)  \\
     \int_{x\in l}  \check f_l^k(x)dx & \int_{x\in h}  \check f_h^k(x)dx
    \end{bmatrix}$$ to be full rank. Note that the determinant of matrix $M$ can be represented in the following:    
    \begin{eqnarray*}
    det(M)=\check f_l^k(\xL)\int_{x\in h}  \check f_h^k(x)dx+ \int_{x\in l}  \check f_l^k(x)dx \check f_h^k(\xL).
    \end{eqnarray*}
As all components are positive, $det(M)>0$, matrix $M$ is full rank. Consequently, the scales can be pinned down in the following closed-form expression:
    \begin{eqnarray*}
    \begin{bmatrix}
    \eta^k_l\\
    \eta^k_h 
    \end{bmatrix} =M^{-1} \begin{bmatrix}
    0\\
    1
    \end{bmatrix}.
    \end{eqnarray*}
    
     Once the scales are pinned down, we can identify state weight $p_k$.

\subsection{Proof of Lemma \ref{lemma_onlyn}} 

Recall that the distribution of the observed $r^{th}$ order statistic is a mixture 
\[
F_r(x) = \sum_n  p_n \sum_{i=r}^{n} c_{i,n} [\eta \check{F}(x)]^{i}[1-\eta \check{F}(x)]^{n-i}, \forall x \le \xL,
\] 
whose left-hand side is known and right-hand side is linear in $p_n$ and polynomial in $\eta$. For any value in $\alpha \in [0,1]$, we can find an $x$ such that $\hat{F}(x) = \alpha$. Therefore, a suitable set $\{\alpha_1, \ldots, \alpha_{|n|}\}$ with the restriction $\sum_n p_n=1$ leads to a system of $|n|+1$ equations that provides restrictions on both $p_n$s and $\eta$. To better understand the  intuition, we can solve the system of linear equations, given the set of $|n|$ different values of $\alpha$,  and obtain $p_n$ as a function of $\eta$. We then plug all $p_n$s into the restriction that $\sum_n p_n=1$, resulting in a scalar polynomial function with $\eta$ as the only unknown. That is, we transform the original problem into a simple problem of pinning down $\eta$ at any set of $x$. The common root of the equation for all  $x\leq \xL$ is the true $\eta$.

Once the distribution of competition $p_{n}$ is identified, we can identify the bid distributions for $y  \ge \xL$ in the following two steps. First, the distribution of the mixture of competition is identified from Equation \eqref{eq_second_1}, because density $f_{r-1:r-1}(y)$ for $y \le \xL$ is identified given that scale $\eta$ is known. Second, we identify the bid distribution using the fact that the identified mixture component is a monotonic function of the to-be-identified bid distribution. That is, 
\begin{eqnarray}
\int^{\bar x}_y \sum_n p_{n}   \frac{n!}{(r \!-\!1)!(n\!-\! r \!+\!1)!}  f_{1:n \!-\! r\!+\!1}(y) dy = \sum_n p_{n} \frac{n!}{(r \!-\! 2)!(n \!-\! r)!} \left[ 1 \!-\! F(y) \right]^{n \!-\! r \!+\!1},
\label{eq_mon}
\end{eqnarray}
where the left-hand side is identified while the right-hand side is strictly decreasing in unknown function $F(y)$,  $\forall y > \xL$. Therefore, there is a unique solution.

\bigskip

\bigskip

    \newpage 
    
    \bibliographystyle{ecta}
    \bibliography{refOS}

    \end{document}